\input harvmac 
\input epsf.tex

\overfullrule=0mm
\def\IR{\relax{\rm I\kern-.18em R}} 
\font\cmss=cmss10
\font\cmsss=cmss10 at 7pt 
\def\IZ{\relax\ifmmode\mathchoice
{\hbox{\cmss Z\kern-.4em Z}}{\hbox{\cmss Z\kern-.4em Z}}
{\lower.9pt\hbox{\cmsss Z\kern-.4em Z}} 
{\lower1.2pt\hbox{\cmsss Z\kern-.4em Z}}
\else{\cmss Z\kern-.4em Z}\fi}
\def\file#1{#1}
\def\figbox#1#2{\epsfxsize=#1\vcenter{
\epsfbox{\file{#2}}}} 
\newcount\figno
\figno=0
\def\fig#1#2#3{
\par\begingroup\parindent=0pt\leftskip=1cm\rightskip=1cm\parindent=0pt
\baselineskip=11pt
\global\advance\figno by 1
\midinsert
\epsfxsize=#3
\centerline{\epsfbox{#2}}
\vskip 12pt
{\bf Fig. \the\figno:} #1\par
\endinsert\endgroup\par
}
\def\figlabel#1{\xdef#1{\the\figno}}
\def\encadremath#1{\vbox{\hrule\hbox{\vrule\kern8pt\vbox{\kern8pt
\hbox{$\displaystyle #1$}\kern8pt}
\kern8pt\vrule}\hrule}}


\font\cmss=cmss10 \font\cmsss=cmss10 at 7pt
\def\IZ{\relax\ifmmode\mathchoice
{\hbox{\cmss Z\kern-.4em Z}}{\hbox{\cmss Z\kern-.4em Z}}
{\lower.9pt\hbox{\cmsss Z\kern-.4em Z}}
{\lower1.2pt\hbox{\cmsss Z\kern-.4em Z}}\else{\cmss Z\kern-.4em Z}\fi}

\Title{\vbox{\hsize=3.truecm \hbox{SPhT/99-073, NBI-HE-99-22}}}
{{\vbox {
\bigskip
\centerline{Integrable 2D Lorentzian Gravity}
\medskip
\centerline{and Random Walks}
}}}
\bigskip
\centerline{P. Di Francesco\foot{philippe@spht.saclay.cea.fr, partially 
supported by NSF
grant PHY-9722060},
E. Guitter\foot{guitter@spht.saclay.cea.fr}}
\medskip
\centerline{ \it CEA-Saclay, Service de Physique Th\'eorique,}
\centerline{ \it F-91191 Gif sur Yvette Cedex, France}
\medskip
\centerline{C. Kristjansen\foot{kristjan@alf.nbi.dk, supported by
the Carlsberg Foundation}}
\medskip
\centerline{ \it The Niels Bohr Institute,}
\centerline{ \it Blegdamsvej 17, DK-2100 Copenhagen \O, Denmark}
\medskip


\noindent
We introduce and solve a family of 
discrete models of 2D Lorentzian gravity with higher 
curvature weight, which
 possess mutually commuting transfer matrices, and whose 
spectral parameter interpolates between flat and curved space-times. We 
further establish a one-to-one correspondence between Lorentzian triangulations 
and directed Random Walks. This gives a simple explanation why the Lorentzian
triangulations have fractal dimension $2$ and why the curvature model lies in 
the universality class of pure Lorentzian gravity. We also study integrable 
generalizations of the curvature model with arbitrary polygonal tiles. All of 
them are found to lie in the same universality class.

\noindent

\Date{07/99}


\nref\KKMW{H.\ Kawai, N.\ Kawamoto, T.\ Mogami, and Y.\ Watabiki,
Phys.Lett. B306 (1993) 19, hep-th/9302133}
\nref\Wat{Y.\ Watabiki, Nucl. Phys. B441 (1995) 119, hep-th/9401096}
\nref\AW{J.\ Ambj\o rn, and Y.\ Watabiki, Nucl.Phys. B445 (1995) 129-144,
hep-th/9501049}
\nref\AJW{J. Ambj\o rn, J. Jurkiewicz and Y. Watabiki, 
Nucl. Phys. B454 (1995) 313, hep-th/9507014}
\nref\AAMT{J. Ambj\o rn, K.N. Anagnostopoulos, U. Magnea and
G. Thorleifsson, Phys.Lett. B388 (1996) 713-719, hep-lat/9606012}
\nref\AA{J.\ Ambj\o rn and A.N.\ Anagnostopoulos, Nucl.Phys. B497
(1997) 445-478, hep-lat/9701006}
\nref\ABJ{J. Ambj\o rn, P. Bialas and J. Jurkiewicz, JHEP 9902 (1999) 005,
hep-lat/9812015}
\nref\IK{N.\ Ishibashi and H. Kawai, Phys.\ lett.\ B322 (1994) 67, 
hep-th/9312047, Phys.\ Lett.\ B352 (1995) 75, hep-th/9503134}
\nref\AKW{J.\ Ambj\o rn, C.\ Kristjansen and Y.\ Watabiki, 
Nucl.Phys. B504 (1997) 555-579, hep-th/9705202}
\nref\AAJK{J.\ Ambj\o rn, K.N.\ Anagnostopoulos, J.\ Jurkiewicz and
C.\ Kristjansen, JHEP 9804 (1998) 016, hep-th/9802020}
\nref\GK{S.\ Gubser and I.\ Klebanov, Nucl. Phys. B416 (1994) 827, 
hep-th/9310098}
\nref\AL{J.\ Ambj\o rn and R.\ Loll, Nucl.\ Phys.\ B536 (1998) 407,
hep-th/9805108}
\nref\AMB{J.\ Ambj\o rn, J.\ Nielsen, J.\ Rolf and R.\ Loll, 
Chaos Solitons Fractals 10 (1999) 177-195}
\nref\NA{R.\ Nakayama, Phys.\ Lett.\ B325 (1994) 347, hep-th/9312158}
\nref\AAL{J.\ Ambj\o rn, K.N.\ Anagnostopoulos and R.\ Loll,
hep-th/9904012}
\nref\Apriv{J.\ Ambj\o rn, private communication}
\nref\Wein{S.\ Weinberg, {\it Ultraviolet divergences in quantum
gravity}, in ``General Relativity, an Einstein centenary survey''
edited by S.W.\ Hawking and W.\ Israel, Cambridge University Press 1979}
\nref\Kaz{V.\ A.\ Kazakov, M.\ Staudacher and T.\ Wynter,
Commun.\ Math.\ Phys.\  179 (1996) 235 (hep-th/9506174),
Nucl.\ Phys.\ B471 (1996) 309 (hep-th/9601069)}\nref\Stau{M.\ Staudacher, 
Nucl.\ Phys.\ B336 (1990) 349}
\nref\sinai{Ya.G.\ Sinai, Theor.\ Prob.\ Appl.\ 27 (1982) 256}
\nref\CG{Such expressions for the Sinai model were first derived in 
A.\ O.\ Golosov, Sov.\ Math.\ Dokl.\ 28 (1983) 18. For a more direct derivation
based on a real space renormalization group analysis, see 
P.\ Le Doussal, C.\ Monthus and D.S.\ Fisher, Phys. Rev. E 59 (1999) 4795, 
cond-mat/9811300.
A derivation based on discrete random walks can be found in 
J.\ Chave and E.\ Guitter, J.\ Phys.\ A: Math.\ Gen.\ 32 
(1999) 445, cond-mat/9809087}



\newsec{Introduction}
The invention of a transfer matrix technique for dynamical
triangulations [\xref\KKMW,\xref\Wat] has made it possible to introduce a
concept of distance in two-dimensional quantum gravity. This again has
made it possible, in the case of pure gravity,
to calculate analytically a loop-loop correlation
function as a function of geodesic distance and to show that the
scaling relations from flat space statistical mechanics have analogues
in the quantum theory \AW. An
intriguing result of these investigations is that the fractal
dimension of the quantum space times turns out to be four and not
two. Unfortunately, when it comes to the inclusion of matter fields
the transfer matrix technique for dynamical triangulations has not yet
proven as efficient as its regular lattice counterpart. Correlators as
a function of geodesic distance have been studied numerically 
[\xref\AJW--\xref\ABJ] but
analytical calculations have not been possible.\foot{There have,
however, been such calculations using an alternative, fully valid, but
not equivalent definition of distance [\xref\IK--\xref\AAJK], 
see also \GK .}

\fig{A typical discretized universe of $(1+1)$-dimensional
Lorentzian quantum gravity. Each constant-time strip is an
arbitrary succession of triangles pointing up or down. Boundary conditions
can be chosen to be periodic, free, fixed according to the model at
hand.}{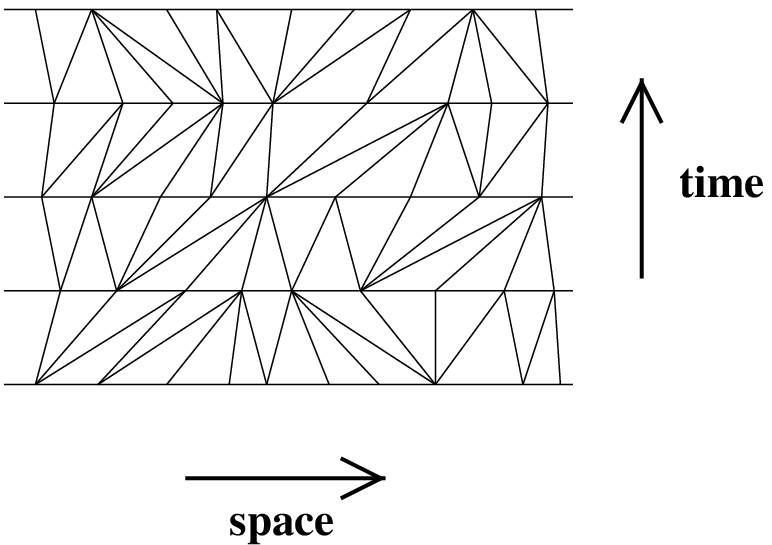}{8.cm}
\figlabel\lorgra

Recently, a different discrete approach to two-dimensional quantum
gravity has been proposed [\xref\AL,\xref\AMB]. In this approach which is known
as Lorentzian quantum gravity only triangulations which admit a causal
structure are allowed in the state sum. 
An example of such a
triangulation is shown in Fig.\ \lorgra. The elementary building blocks
are triangles with one space-like and two time-like edges which are
glued together at random so as to form a piling of constant-time slices.
In a sense such triangulations represent only the space fluctuations of the
universe while respecting chronology. In Lorentzian quantum gravity,
baby universes are not allowed. In other words, a given time slice of
the triangulation has only one connected component. This, in
particular, means that a universe always has spherical topology. 
The
model of pure Lorentzian quantum gravity was solved in \AL\ exploiting the
fact that the generating functional for its transfer matrix obeyed a
certain iterative equation. The continuum limit of the model proved to
be different from that of usual Euclidean quantum gravity or Liouville
theory. For instance, the fractal dimension of the quantum universes
turned out to be equal to two. However, the continuum limit of
Lorentzian quantum gravity is in accordance with continuum results
obtained from calculations carried out in the proper time gauge
\NA.\foot{If 
baby universes are allowed in Lorentzian quantum gravity the
usual Euclidean gravity continuum limit is obtained. We refer to \AL\
for a discussion of this point.}
A Lorentzian lattice by construction is something between a regular
lattice and a truly random one. It is therefore interesting to know if
the introduction of matter fields on such lattices leads to any
interaction between matter and geometry. In \AAL\ the effect of
coupling an Ising model to Lorentzian quantum gravity was
studied. The investigations were based partly on a high temperature
expansion, partly on Monte Carlo simulations. No sign of interaction between
matter and geometry was seen. Numerical simulations currently being carried 
out indicate, however, that for matter fields with $c>1$ a non-trivial
interaction between matter and geometry takes place \Apriv. 

In a first part of the present article we formulate and solve
exactly various other extensions of two-dimensional Lorentzian
quantum gravity.
The simplest extension, which we develop in details throughout 
the paper, is that of Lorentzian gravity with a higher curvature term 
which is equivalent to 
Lorentzian quantum gravity coupled to a simple dimer model or to
Lorentzian quantum gravity with universes built from triangles and
squares. We show that this model is integrable in the sense that its
transfer matrix can be explicitly diagonalized and that there exists a
set of commuting transfer matrices which allow us to interpolate
between regular and random Lorentzian lattices. Furthermore we
describe how to construct models of Lorentzian quantum gravity
allowing general $p$-gons as building blocks while preserving the
integrability structure. As will become evident the transfer matrix
formulation is by far more efficient for Lorentzian than for
Euclidean triangulations. In fact the integrability structure revealed
has many treats in common with the integrability structure found in
regular lattice transfer matrix studies and it is our hope that our
further investigation of this structure will enable us to solve
exactly more realistic systems of matter fields coupled to Lorentzian
quantum gravity such as Lorentzian triangulations equipped with Ising
spins. For all the models that we have considered so far the
universality class of the geometrical system  is the same as that of
pure Lorentzian quantum gravity.

In a second part of this article we study the relationship between
Lorentzian quantum gravity and random walks. We prove that there
exists a one-to-one correspondence between Lorentzian triangulations
and directed random walks drawn on the regular triangular lattice. This allows
us to set up a dictionary connecting concepts in Lorentzian quantum
gravity to concepts in the theory of random walks. For instance, 
the expression for the loop-loop correlation function follows from a similar
expression for large excursion probabilities for random walks. Furthermore,
the integrability structure of the model of Lorentzian quantum gravity
with a higher curvature term can be understood in terms of the possibility of
a simple block decomposition of the corresponding random walk.
Finally the
random walk equivalence provides an explanation why Lorentzian
triangulations have fractal dimension two and  why we can not obtain
from our model with a higher curvature term other critical behaviour than
that of pure Lorentzian quantum gravity.\foot{Dimensional analysis
indicates that adding a higher curvature term to the
Einstein-Hilbert action should not modify continuum physics. However,
this argument is only valid perturbatively and
one could still hope for the existence of a (non-perturbative)
ultra-violet fixed point, a scenario (in 4D) denoted by Weinberg as asymptotic
safety \Wein.}

The simple extension of Lorentzian quantum gravity involving
a higher curvature term as well as its various alternative interpretations
are described in Sect. 2. Sect. 3 treats the
equivalence between random walks and Lorentzian triangulations and
Sect. 4 contains a discussion of Lorentzian gravity involving
general polygonal building blocks. Finally, in Sect. 5 we
conclude and discuss the future prospects of transfer matrix
techniques for Lorentzian triangulations.

\newsec{Discrete Lorentzian 2D Gravity via Triangulations } 

\subsec{Lorentzian Quantum Gravity with a Higher Curvature Term}

We consider quantum universes of the type depicted in Fig.\ \lorgra.
Generally we will be interested in a time evolution from, say, time
$t=1$ to $t=T$, i.e.\ a time strip of width $T$. In the course of the
article we will consider various possible boundary conditions in
space. These will be explained at the relevant points.

As usual in two-dimensional gravity, we will attach a weight $g$ per triangular
face, resulting in an overall factor of $g^A$ for each triangulation, where
$A$ is the total number of triangles, measuring the area if we decide that
triangles have unit area. 

We now generalize the model by also attaching to each triangulation 
an intrinsic curvature weight defined as follows. 
For a two-dimensional triangulated manifold, curvature resides on
vertices, and the curvature at a vertex is proportional to $(v-6)$
where $v$ is the valence of the vertex. Here we do not wish to add an
ordinary curvature term (which is trivial in two dimensions) but a
term which suppresses (or enforces) local curvature. Due to the time
slice structure of the triangulation we take as a measure of the
curvature at a vertex the quantity $|v_1-3|+|v_2-3|$ where $v_1$ and
$v_2$ are the number of triangles adjacent to the vertex in the upper
and the lower time-slice respectively, and we attach to each vertex of
the triangulation the weight
$a^{(\vert v_1-3\vert +\vert v_2-3\vert)/2}$. Introducing this weight
factor can be viewed as adding a higher curvature term to the
Einstein-Hilbert action (note, however, that one can have $v-6=0$ and
$|v_1-3|+|v_2-3|\neq 0$). Our model is similar in spirit to the
models considered in \Kaz, where the effect of adding a higher
curvature term to usual dynamically triangulated gravity was
considered.
Since any vertex is linked to exactly two up (resp.\ down) triangles is its 
upper (resp.\ lower) slice, with an arbitrary number of down (resp.\ up) 
triangles in-between, a simpler, completely equivalent way to introduce the
curvature weight is to attach a weight $a$ per {\it pair} of adjacent 
triangles pointing both up or both down, within the same constant-time slice.
The graphical representation of Fig.\lorgra\ is more transparent 
in the dual picture, where a constant-time strip becomes a succession of
half-edges attached to the dual constant-time line, where the
half-edges lying above (resp.\ below)
the line correspond to triangles pointing down (resp.\ up).
Moreover, as each triangle has exactly one space-like edge, the triangles
pointing down in the strip of time $t$ all share an edge with the triangles
pointing up in the strip of time $t+1$, henceforth the dual half-edges are 
connected to form edges between the lines of time $t$ and $t+1$.
\fig{The Boltzmann weights of the Lorentzian gravity model
with a higher curvature term. Each vertex (of the dual lattice) receives a
factor $g$. In addition each sequence of neighbouring 
up-up or down-down half-edges
receives a factor $a$. We have represented a typical world-sheet 
configuration.}{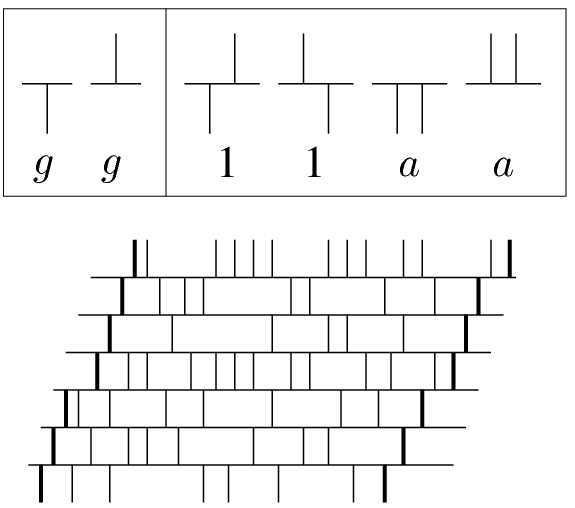}{9.5cm}
\figlabel\boltz
The Boltzmann weights defined above 
are depicted in Fig.\boltz, where we also
give an example of a world-sheet configuration in the dual picture.

The Lorentzian gravity problem above, and its generalization with
intrinsic curvature admit a transfer matrix formulation. 
Indeed, if we compute the Boltzmann weight of a single time-line with $i$ 
lower half-edges and $j$ upper ones, and sum over all possible relative
positions of these half-edges, we find a number $T_{i,j}(g,a)$
that depends only on the numbers of half-edges, but not on their specific
relative positions. This can in turn be taken as the transfer matrix element
$(i,j)$, that transfers from a row of $i$ half-edges to one of $j$.
For instance, the partition function of a strip of time width $T$,
with $\ell_1$ lower triangles pointing up and $\ell_2$ upper ones pointing down
is simply
\eqn\transmat{ Z_{\ell_1,\ell_2}(T\vert g,a)~
=~ \bigg(T(g,a)^T\bigg)_{\ell_1,\ell_2}.}
For periodic boundary conditions, this is the so-called two-loop correlator
of Lorentzian gravity, which describes world sheets of cylindric form bounded
by two loops of respective lengths $\ell_1$ and $\ell_2$. 

Let us  adopt the following choice of 
fixed boundary conditions,
that differ from the periodic ones of \AMB. 
We assume that each time strip always has one lower half-edge 
(i.e. triangle pointing up) on its 
leftmost end and one upper one (triangle pointing down) on its rightmost end. 
This implies in particular that $i,j\geq 1$ in the transfer matrix, i.e. that
the time slices never degenerate into the vacuum.
These boundary conditions can be described as ``staircase'' conditions 
since they imply that the left and right hand side space-like boundaries of 
the world-sheet
have the shape of a staircase with the stairs extending
to the right at each successive time line (see the example of
Fig.\boltz). 
For reasons that will become clear later, we attach a weight $\sqrt{g}$ only
to each of the two {\it boundary} half-edges in each slice. 

We may now compute the transfer matrix element $T_{i,j}(g,a)$ by
summing over all possible configurations of upper and lower half-edges.
This is done by first summing over the number $k\geq 1$ of blocks made,
say, of $n_r\geq 1$ consecutive lower half-edges, followed by $m_r\geq 1$
upper  ones, $r=1,2,...,k$, and then summing over the corresponding
partitions of $i=\sum_r n_r$ and $j=\sum_r m_r$, with positive integers 
$n_r,m_r$. 
Note that the conditions $n_1\geq 1$ 
and $m_k\geq 1$ are a direct consequence of the boundary conditions 
we have imposed.  
We get
\eqn\weget{ \eqalign{
T_{i,j}(g,a)&= \sum_{k\geq 1} \sum_{n_r,m_r\geq 1,\ r=1,2,...,k\atop
\Sigma n_r=i,\ \Sigma m_r=j} \figbox{5.cm}{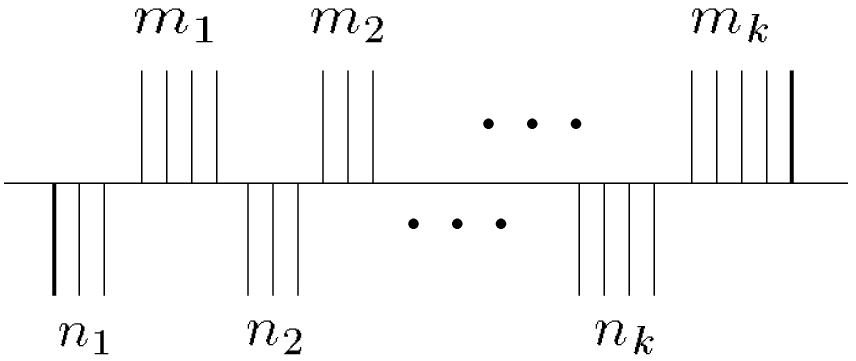} \cr
&= \sum_{k\geq 1} \sum_{n_r,m_r\geq 1,\ r=1,2,...,k\atop
\Sigma n_r=i,\ \Sigma m_r=j}
g^{i+j-1} a^{\Sigma (n_r-1)+\Sigma (m_r-1)} \cr
&=g^{i+j-1} a^{i+j} \sum_{k\geq 1} {1\over a^{2k}} {i-1 \choose k-1}
{j-1 \choose k-1}. \cr}}
This is in general a polynomial of $a$, as the range of the summation 
is finite ($k\leq \min(i,j)$).

The formula \weget\ is best encapsulated into the generating function
\eqn\generat{\eqalign{
\theta(x,y\vert g,a)&=\sum_{i,j\geq 1} x^i y^j T_{i,j}(g,a)\cr
&={1\over g}\sum_{k\geq 1} {1\over a^{2k}} \sum_{i\geq k} {i-1 \choose k-1}
(gax)^i \sum_{j\geq k} {j-1 \choose k-1} (gay)^j \cr
&= {1\over g}\sum_{k\geq 1} {1\over a^{2k}} 
{(gax)^k (gay)^k\over (1-gax)^k (1-gay)^k}\cr
&={g xy\over 1-ga(x+y)-g^2(1-a^2)xy}.\cr}}

In the particular case $a=1$, this reduces to the generating function 
for pure Lorentzian gravity discretized via triangulations to be compared
with that of \AL, but with different boundary conditions.
When $a=0$, \generat\ reduces to the generating function for flat
world-sheets. Indeed, $\theta(x,y\vert g,0)=gxy/(1-g^2xy)$ simply
means that $T_{i,j}(g,0)=g^{i+j-1}\delta_{i,j}$. In other words,
the pattern of triangles becomes regular (successions of triangles pointing
up,down,up,...etc...), and the discrete world-sheet becomes a rhombus-shaped
portion of the triangular lattice, with area $2 \ell \times T$, where
$\ell$ denotes the number of lower triangles pointing up in the first
time slice.
Hence our parameter $a$ allows us to interpolate between flat and curved
spaces. 

\subsec{Diagonalization of the transfer matrix}

To solve our model, and be able to compute quantities like \transmat,
the simplest strategy is to diagonalize the transfer matrix $T(g,a)$.
But as the number of triangles in each time slice is not bounded,
the matrix has infinite size. Nevertheless, at the expense of extra
precautions, we have been able to construct a complete set of orthonormal
eigenvectors for $T(g,a)$, as a linear map acting on the 
infinite-dimensional vector space $E=\IR \otimes \IR \otimes ... $. 
They read as follows. 

Unless otherwise noted, we will always assume that the parameters
$g$ and $a$ are real and satisfy the inequalities
\eqn\ineq{ \vert {1-g^2(1-a^2)\over ga} \vert >2, \qquad |ga|<1.}
Let $q\equiv q(g,a)$ be the real solution to the
quadratic equation
\eqn\quadrat{ ga (q+{1\over q}) =1-g^2(1-a^2), }
such that $|q|<1$. Then the functions
\eqn\genevec{\eqalign{
F_m(x\vert q)&= \sum_{i\geq 1} x^i v_i^{(m)}(q)\cr
&= \sqrt{1-q^2}\; {x(q-x)^{m-1}\over (1-qx)^{m}},\cr}}
for $m=1,2,3,...$ are the generating functions for the components
$v_i^{(m)}$ of the $m$-th eigenvector $v^{(m)}\equiv v^{(m)}(g,a)$ 
for $T(g,a)$.  Moreover
the corresponding vectors form an orthonormal basis of $E$. 

The second statement is readily proved by considering the following
contour integral over the unit circle
\eqn\orton{\eqalign{ 
v^{(m+p)}\cdot v^{(m)}&=\sum_{i\geq 1} v_i^{(m+p)} v_i^{(m)}\cr
&=\oint {dx \over 2 i \pi x} F_{m+p}(x\vert q) F_{m}({1\over x}\vert q)\cr
&=-(1-q^2)\oint {dx \over 2i\pi} {(q-x)^{p-1}\over (1-qx)^{p+1}}\cr
&=\left\{ \matrix{ 0 & {\rm if} \ \ p\geq 1\cr
1 & {\rm if} \ \ p=0\cr} \right. \cr}}
where, using the Cauchy residue formula, we have noted that there
was no pole of the integrand inside the unit disc when $p\geq 1$
(recall that $|q|<1$),
and the result for $p=0$ is simply given by the residue at the pole $x=q$. 
This proves the orthonormality of the set of 
vectors $\{v^{(m)}\}_{m=1,2,...}$.
It is also easy to see that the matrix $V$ with entries 
$V_{i,m}=v_i^{(m)}$ is symmetric. This is proved by noting that  
the generating function 
\eqn\genevect{v(x,y\vert q)=\sum_{i,m\geq 1}
x^i y^m v_i^{(m)}={\sqrt{1-q^2}\, xy \over 1-q(x+y) +xy }, }
is manifestly symmetric in $x$ and $y$. From this 
symmetry and the orthogonality relation \orton ,
we deduce the following completeness relation
\eqn\compl{\sum_{m\geq 1} v_i^{(m)} v_j^{(m)} = \delta_{ij}\ .}

The first statement above, that $v^{(m)}$ be the $m$-th eigenvector of 
$T(g,a)$, is proved analogously, by means of a contour integral over
the unit circle:
\eqn\prevec{\eqalign{\sum_{i\geq 1} x^i \bigg(
&T(g,a)v^{(m)}\bigg)_i  = \sum_{i,j\geq 1}x^i T_{i,j}(g,a)
v_j^{(m)} \cr
&=\oint {dy \over 2i\pi y} \theta(x,y\vert g,a) 
F_m({1\over y}\vert q) \cr
&=\sqrt{1-q^2}
\oint {dy \over 2i\pi y} {g xy \over (1-gax)-y(ga+g^2(1-a^2)x)}
{{1\over y}(q-{1\over y})^{m-1}\over (1-{q\over y})^{m}}\cr
&={g x\over ga +g^2(1-a^2)x} 
F_m\big({ga+g^2(1-a^2)x\over 1-gax}\vert q\big)\cr
&=\sqrt{1-q^2} g x {\big((1-gax)q -(ga+g^2(1-a^2)x)\big)^{m-1} \over
\big(1-gax-q(ga+g^2(1-a^2)x)\big)^{m}}.\cr}}
Note that only the pole $y=y_0(x)=(1-gax)/(ga+g^2(1-a^2)x)$ has contributed
to the contour integral, as we have taken now only the poles lying 
{\it outside}
the unit disc, and $y_0(x)>1$ for $|x|<1$, thanks to the inequalities 
\ineq. 
Using the equation \quadrat, we may write $g^2(1-a^2)=1-ag(q+{1\over q})$,
and notice that the numerator and denominator monomials in the last
line of \prevec\ respectively read
\eqn\resred{\eqalign{
(1-gax)q -(ga+g^2(1-a^2)x) &= (q-x)(1-{ga\over q}),\cr
1-gax-q(ga+g^2(1-a^2)x) &= (1-qx)(1-q g a),\cr}}
so that we finally arrive at the eigenvalue equation
\eqn\eigenvecs{
T(g,a)v^{(m)} = \Lambda_m v^{(m)}, }
where
\eqn\eigenval{ \Lambda_m\equiv \Lambda_m(g,a)= g{ (1-{ga\over q})^{m-1}
\over (1-qga)^{m}}, }
for $m=1,2,3,...$

Let us examine 
our result in the pure Lorentzian
gravity case $a=1$. The quadratic equation \quadrat\ is solved as 
\eqn\solq{q(g,1)=gC(g^2),}
for $\vert g\vert <{1\over 2}$,
where
\eqn\catal{ C(x)= {1-\sqrt{1-4x}\over 2x}= 
\sum_{n\geq 0} {(2n)!\over (n+1)! n!} x^n,}
is the generating function of the Catalan numbers, satisfying
$xC^2(x)=C(x)-1$.
The eigenvalues can then be simplified to read
\eqn\evalm{\Lambda_m=[gC(g^2)]^{2m-1}=q^{2m-1}.} 
Note that as $|q|<1$, we have $\Lambda_1>\Lambda_2>...>0$, and
all the eigenvalues have series expansions in powers of $g$ with positive 
integer coefficients.

\subsec{Commuting transfer matrices}

We now make the crucial observation that the eigenvectors $v^{(m)}(g,a)$
of the previous section only depend on $q$, which is itself a certain
function of $g$ and $a$. Hence there exists an infinite family
of matrices $T(g,a)$ sharing the same eigenvectors, namely those
for which the values of $g,a$ lead to the same value of $q$.   
But matrices which can be diagonalized simultaneously form a commuting
set.
More precisely, the following statement holds:
\eqn\stat{ [T(g,a),T(g',a')]=0 \ \ \ {\rm iff} \ \ \ 
{1-g^2(1-a^2)\over ga } = {1-(g')^2(1-(a')^2)\over g'a'} = q+{1\over q}, }
for arbitrary $q$ (assumed to be real and such that $\vert q\vert <1$).
Moreover, the common eigenvectors are given by \genevec.
Again, Eq.\stat\ can be proved directly by use of contour integrals
involving the product of two generating functions \generat. 

To better understand the mechanism of this commutation, let us 
express the transfer matrix $T$ in terms of its
orthonormal eigenvectors and of its eigenvalues, namely
\eqn\eveceval{\eqalign{
\theta(x,y\vert g,a)~&=~ \sum_{m\geq 1} F_m(x|q)
\Lambda_m F_m(y|q) \cr
&=~ {g \over 1-qga}\sum_{m\geq 1}F_m(x|q) 
\lambda^{m-1} F_m(y|q) \cr
&=~ \sqrt{\lambda} \sum_{m\geq 1}F_m(x|q) 
\lambda^{m-1} F_m(y|q), \cr}}
where we have identified
\eqn\poval{\lambda= {1-{ga\over q} \over 1-qga}, }
by use of \eigenval.
For any fixed $q$ with $\vert q\vert <1$, we may use the (spectral) 
parameter $\lambda$ to
characterize the different commuting matrices \stat.
Denoting 
\eqn\nota{ T_q(\lambda)= T(g,a) \ \ \ {\rm for} \ \ 
{1-g^2(1-a^2)\over ga }=q+{1\over q} \ \ {\rm and} \ \ 
ga= {1-\lambda \over {1\over q}-q\lambda},}
and 
\eqn\vpta{ V_q^{(m)}= v^{(m)}(g,a),}
{\it independently} of $\lambda$, we have from \eveceval\ 
\eqn\evecq{T_q(\lambda)=\sqrt{\lambda}\sum_{m\ge 1} \lambda^{m-1} V_q^{(m)}
(V_q^{(m)})^t \ ,}
where we have used the orthogonal projector 
$P_q^{(m)}\equiv V_q^{(m)}(V_q^{(m)})^t$ 
onto the $m$-th eigenspace of $T_q(\lambda)$. 
Thanks to the orthogonality relation for the eigenvectors, which translates
into $P_q^{(m)} P_q^{(m')}=\delta_{m,m'} P_q^{(m)}$, we easily get that
\eqn\addi{ T_q(\lambda) T_q(\lambda') 
= T_q(\lambda \lambda'). }
This relation trivially implies the commutation of transfer matrices
\stat. But this is a much stronger constraint. Note that all real 
values of $\lambda$ and $q$ such that $|q|<1$ are allowed.  
Note also that both the commutation \stat\ and the multiplicativity
property of spectral parameters \addi\ hold only with 
the help of the curvature weight $a$. The ``solvability'' of the case
without a curvature term (allowing for the solution of \AL) is
simply a consequence of these more general properties.
Finally, let us mention that the transfer matrix of the regular lattice
is a member of the family indexed by the parameter $q$ for {\it any} $q$,
as it can be obtained by taking the limit $a\to 0$, $g\to 1$ while 
keeping the ratio $(1-g^2(1-a^2))/(ga)=q+(1/q)$ fixed.
In this limiting case, $\lambda=1$ and the transfer matrix 
$T_q(1)$ is nothing but the identity matrix.

To conclude this section, let us rewrite the generating function 
\generat\ in terms of $q$ and $\lambda$ only
\eqn\ragen{\eqalign{
\theta_q(x,y\vert \lambda) &= \sum_{i,j\geq 1}
x^i y^j \big[T_q(\lambda)\big]_{i,j} \cr
&= { \sqrt{\lambda} x y (1-q^2) 
\over (1-qx)(1-qy)- \lambda (q-x)(q-y)}.\cr}}

\subsec{Squares and Triangles}

The curvature model introduced above may be reinterpreted as a discrete
model for Lorentzian gravity, where the world-sheet is generated by arbitrary
tessellations with squares and triangles that respect  chronology.
This is easily seen by performing the following transformation on
the configurations of the previous model.

\fig{The Boltzmann
weights of the square and triangle formulation of the curvature model for
discrete 2D Lorentzian gravity. We have also represented a typical
world-sheet configuration in this new interpretation,
together with its
dual, made of squares and triangles.}{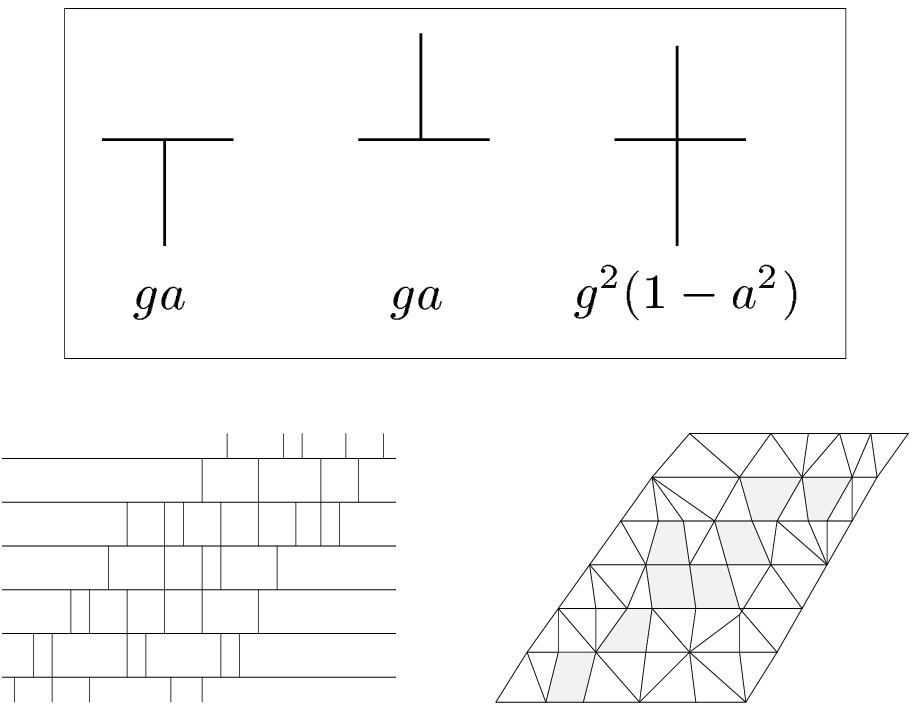}{10.5cm} 
\figlabel\boltzto

Let us expand again in powers of $x$ and $y$ the generating function \generat\
\eqn\genexpand{
\theta(x,y\vert g,a)={g xy\over 1-ga(x+y)-g^2(1-a^2)xy}.}
and interpret each contribution to $x^i y^j$ as a configuration of
$i$ lower and $j$  upper half-edges. By expanding the denominator
of \genexpand , we get an arbitrary sequence of terms $gax$, $gay$
or $g^2(1-a^2)xy$. Whenever a term $gax$ (resp.\ $gay$) is picked,
we interpret it as an isolated lower (resp.\ upper) half edge, which comes
with a weight $ga$. Whenever a term $g^2(1-a^2) xy$ is picked,
we now have a pair of a lower and an upper half-edge which we can
regroup so as to form a crossing of the time line (see Fig.\boltzto). Such
crossings come with a weight factor $g^2(1-a^2)$. Finally, the
numerator $gxy$ in \genexpand\ is there to ensure staircase boundary
conditions for these new configurations. It corresponds
to adding a lower half-edge at the far left and an upper one at the far 
right, each of which comes with a factor $\sqrt{g}$ only (and no $a$).

We thus get another representation for the 
configurations contributing to the transfer matrix element $T_{i,j}(g,a)$,
with the Boltzmann weights defined in Fig.\boltzto. Dually, these correspond
to  supplementing the triangulations we have considered so far, by {\it
squares} with two time-like and two space-like edges (see the example
of a
world-sheet configuration and its dual depicted in Fig.\boltzto). 
All the weights have now been translated into different fugacities
for the triangles and for the squares. Each triangle comes with a weight 
$ga$, while each square receives a weight $g^2(1-a^2)$. Note that this latter 
weight is positive or negative according to whether $a$ is larger or
smaller  
than $1$. Finally, the case of the boundary triangles is special since 
they receive a weight $\sqrt{g}$ only.
Let us also mention that, in the limit $a\to 0$ corresponding to
a regular lattice, only squares and boundary triangles survive, 
in which case it is clear that the only possible arrangement for the 
tessellation is regular and the transfer matrix is the identity matrix 
as it should be.
\fig{The connection between square-triangle tessellations and simple
triangulations is done by splitting each square (i.e. each crossing) 
into two triangles (i.e into two half-edges). To recover the proper
curvature weight, we transfer the $a$ factors as shown by the arrows.
We then add up weights corresponding to the same final triangulation.
}{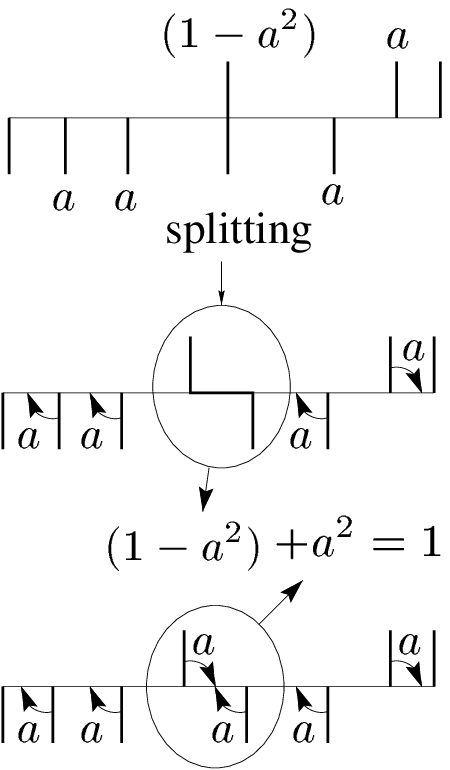}{4.5cm}
\figlabel\split
To make the contact with the curvature model, the idea is to split 
each square of the tessellation into a pair of triangles. To recover
the proper curvature weights, we chose to split each square into a 
down triangle followed by an up one. Now the same triangulation is
obtained from several square-triangle tessellations and the matrix element 
$T_{i,j}(g,a)$ is obtained by summing over all these configurations.
In the dual language, our splitting corresponds to separating each crossing 
into a lower and an upper half, such that the upper-half sits at the 
{\it left} of the lower one (see Fig.\split ). As far as the $g$ factors 
are concerned, since each crossing receives a factor $g^2$, which after 
splitting, gives a weight $g$ per half edge, it is clear that these factors 
are well taken into account. The $\sqrt{g}$ factors at the boundary are
also correct. The $a$ factors are more subtle. Before we collect all 
these factors, we remark that we can transfer the weight $a$ of each 
lower half-edge, which is not at the left boundary, to the horizontal
link sitting just above it and to its left (see Fig.\split ). 
Similarly, we transfer the factor $a$ attached to each upper half-edge, 
which is not at the right boundary, to the horizontal link sitting just 
below to its right. Then all the horizontal links which separate two 
lower half-edges (resp.\ two upper half-edges) receive a factor $a$ from
their right lower half-edge (resp.\ their left upper half-edge) as they 
should in the curvature language, while all the horizontal links with a 
lower half-edge on the left and an upper half-edge on the right receive 
a factor $1$ as they should. Finally the horizontal links with an upper 
half-edge on the left and a lower half-edge on the right receive a weight 
$a^2$ from the two half-edges, together with a contribution $(1-a^2)$ coming
from the configuration where these two half-edges were connected to form 
a crossing. Adding these weights restores a factor $1$, as required.

\subsec{Dimers}

  From our experience with two-dimensional Euclidean quantum gravity we
know that the simplest way of introducing matter in two-dimensional
universes is to allow for dimers. Dimers are decorated links which
have a certain weight associated with them and each triangle in a
triangulation can carry at most one dimer. It is well known that by
fine-tuning the weight of the dimers in two-dimensional Euclidean
quantum gravity to some specific negative value one can reach a
critical point where the universality class of the underlying
geometrical system changes from that of pure gravity to that
characteristic of gravity coupled to non-unitary matter with central
charge $c=-22/5$. It is therefore interesting to note that our model
of triangles and squares can be viewed as a simple dimer model. This
interpretation simply comes about if one views each square as
consisting of two triangles sharing a dimer which lies along either
one of the diagonals of the square. In this picture the triangle
building blocks do not have any dimers associated with them. The weight
of a square, $g^2(1-a^2)$ is hence to be understood as $2\cdot
(ga)^2\big((1-a^2)/(2a^2)\big)$ where $ga$ is the weight of a
triangle, $(1-a^2)/(2a^2)$ is
the weight of a dimer and the factor 2 in front takes into account that
there are two ways of placing the dimer inside the square. From this
decomposition it follows that if new critical behaviour occurs it
should happen at a value of $a$ for which $a>1$. However, it is
important to note that our model only includes a subset of
configurations of the full dimer model since we do not have any dimers
on the space-like links. Thus, if we do not see any new critical
behaviour for the model in question it does not necessarily imply that such
behaviour does not occur for the full dimer model. In this
connection, let us mention that there have been studies of
two-dimensional Euclidean quantum gravity coupled to dimers in a 
spirit similar
to this one. Namely, one has calculated loop-loop correlation
functions as a function of distance, not leaving out any matter
configurations but modifying the concept of distance so as to avoid
dimers on the entrance and exit loop.  In that approach a continuum
behaviour different from that of pure gravity was seen \GK.

\subsec{Solution and continuum limit}

The property \addi\ yields immediately the generating function
of the $T$-th power of the transfer matrix, describing 
a world-sheet of time-size $T$:
\eqn\tetn{\eqalign{ \theta_{q,T}(x,y\vert \lambda)&=
\sum_{i,j\geq 1} x^i y^j Z_{i,j}(T\vert g,a)\cr
&=\sum_{i,j\geq 1} x^i y^j \big[ T_q(\lambda)^T\big]_{i,j}\cr
&= {x y (1-q^2) \lambda^{T\over 2}
\over (1-qx)(1-qy)- \lambda^T (q-x)(q-y)}.\cr}}
for all $T=0,1,2,...$.

As noted in \AMB, one can derive a continuum limit of the
expression \tetn, that corresponds in our notations to
the limit $q\to 1$ ($ga \to a/(a+1)$). 
The critical values $x=y=1$ (corresponding to
the limiting convergence radii of the series $F_m(x|q)$ and
$F_m(y|q)$) must be approached simultaneously. We see then from
\tetn\ that $\lambda$ must also tend to $1$ for the expression
to remain finite. But from \poval, a natural way to realize the
latter limit is to simply keep $a$ constant while $q\to 1$
(and $g\to 1/(a+1)$).

Let us therefore perform the following scaling transformations
for some small parameter $\alpha$ 
\eqn\scatran{ T={\tau\over \alpha}, \qquad x=1-\alpha X, \qquad 
y=1-\alpha Y, \qquad q=e^{-\alpha\sqrt{\Lambda}},}
where $\Lambda$ is the renormalized cosmological constant, $\tau$
the continuous time variable, and
$\alpha$ an infinitesimal parameter with the dimension of a length. 
This corresponds in turn to the scaling behaviour 
\eqn\crilim{
\lambda=1-2a\alpha \sqrt{\Lambda}+{\cal O}(\alpha^2), \qquad
 g={1\over a+1}(1-{a\over 2} \alpha^2 \Lambda) +{\cal O}(\alpha^4),}
for {\it fixed} $a$. 
Applying the transformations \scatran\ to \tetn, 
we get the rescaled two-loop generating function
\eqn\contet{\eqalign{ \Theta(X,Y\vert \tau,\sqrt{\Lambda},a)&\equiv
\lim_{\alpha\to 0} {\alpha} 
\theta_{q,T}(x,y\vert \lambda)\cr
&= {2 \sqrt{\Lambda} e^{-\tau a\sqrt{\Lambda}} 
\over (X+\sqrt{\Lambda})(Y+\sqrt{\Lambda})-e^{-2 \tau a\sqrt{\Lambda}}
(X-\sqrt{\Lambda})(Y-\sqrt{\Lambda})}, \cr}  }
 
The actual two-loop correlator 
\eqn\contlp{G(L_1,L_2\vert \tau,\sqrt{\Lambda},a)  
\equiv \lim_{\alpha\to 0} {1\over \alpha} Z_{l_1,l_2}(T\vert g,a),}
with $L_i=\alpha l_i$, $i=1,2$
is obtained by performing the inverse Laplace
transformation of \contet, leading to (we have set
$\phi=e^{-a\tau\sqrt{\Lambda}}$ for  simplicity)
\eqn\lapinv{ \eqalign{
G&(L_1,L_2\vert \tau,\sqrt{\Lambda},a) = \int_{-i\infty}^{i\infty} 
dXdY e^{L_1X+L_2Y} \Theta(X,Y\vert \tau,\sqrt{\Lambda},a) \cr 
&=\int_{-i\infty}^{i\infty} du dv 
{2\sqrt{\Lambda}\phi e^{\sqrt{\Lambda}(L_1u+L_2v)}
\over (u+1)(v+1)-\phi^2(u-1)(v-1)}\cr
&=2\sqrt{\Lambda}\phi\int_{-i\infty}^{i\infty} du  {e^{\sqrt{\Lambda}
\left(L_1u- L_2{u(1+\phi^2)+(1-\phi^2)\over u(1-\phi^2)+(1+\phi^2)}\right)} 
\over u(1-\phi^2)+(1+\phi^2)} \cr
&= 2\sqrt{\Lambda}\phi e^{-\sqrt{\Lambda}L_2 {1+\phi^2\over 1-\phi^2}}
\int_{-i\infty}^{i\infty} du {e^{\sqrt{\Lambda}\left(
L_1u+L_2 {4\phi^2\over 1-\phi^2}
{1 \over u(1-\phi^2)+(1+\phi^2)}\right)}\over u(1-\phi^2)+(1+\phi^2)}\cr
&= 2\sqrt{\Lambda}\phi  e^{-\sqrt{\Lambda} L_2 {1+\phi^2\over 1-\phi^2}}
\sum_{k\geq 0}  
{\bigg({4\phi^2\over 1-\phi^2}\sqrt{\Lambda}L_2\bigg)^k \over k!} 
\int_{-i\infty}^{i\infty}  {du \ e^{\sqrt{\Lambda}L_1 u}
\over (u(1-\phi^2)+(1+\phi^2))^{k+1}}\cr
&= {2\sqrt{\Lambda}\phi\over 1-\phi^2}  e^{-\sqrt{\Lambda}
L_2 {1+\phi^2\over 1-\phi^2}}
\sum_{k\geq 0} {1\over (k!)^2} \left(
{4\phi^2\over (1-\phi^2)^2} \sqrt{\Lambda} L_2\right)^k {d^k\over du^k} 
e^{\sqrt{\Lambda}L_1u}
\bigg\vert_{u=-{1+\phi^2\over 1-\phi^2}} \cr
&= {2\sqrt{\Lambda}\phi\over 1-\phi^2}  e^{-\sqrt{\Lambda}(L_1+L_2) 
{1+\phi^2\over 1-\phi^2}} \sum_{k\geq 0} {1\over (k!)^2} \left(
{4\phi^2\over (1-\phi^2)^2} \Lambda L_1 L_2\right)^k,\cr}}
where we have first made the change of
variables $X=u\sqrt{\Lambda}$ and $Y=v\sqrt{\Lambda}$, and then used the
Cauchy formula to express the integral as a sum over residues (only one pole
in $v$ contributes, located at 
$v(u)=-(u(1+\phi^2)+(1-\phi^2))/(u(1-\phi^2)+(1+\phi^2))$, whereas only the
multiple poles at $u=(1+\phi^2)/(\phi^2-1)$ occur). In terms of
the rescaled variables, the two-loop correlator reads finally
\eqn\tolop{\eqalign{
G(L_1,L_2&\vert \tau,\sqrt{\Lambda},a)\cr
&= {\sqrt{\Lambda} \over {\rm
sinh}(a\tau \sqrt{\Lambda})} e^{-\sqrt{\Lambda}(L_1+L_2) {\rm
cotanh}(a\tau\sqrt{\Lambda})}  I_0\left({2\sqrt{\Lambda L_1L_2} 
\over {\rm sinh}(a\tau\sqrt{\Lambda})}\right),\cr}} 
where $I_0$ is the modified Bessel function $I_0(x)=\sum_{k\geq 0}
(x/2)^{2k}/(k!)^2$.

A few remarks are in order. First note that the dependence on the curvature 
parameter $a$ is quite simple, and that the physics of the model is not 
affected by it. So we must draw the rather negative conclusion that the 
introduction of curvature in the model cannot change its scaling behaviour. 
It has just shifted the critical value of $g$ (from $1/2$
for the pure gravity case to $1/(a+1)$ in the model with curvature).
Moreover the explicit dependence on $a$ can be entirely absorbed in 
redefinitions of $L_1,L_2,\Lambda$ and $G$, namely 
$L_1\to aL_1$, $L_2\to aL_2$, $\Lambda\to \Lambda/a^2$ and 
$G\to a G$. This ``triviality'' of the curvature 
dependence will be explained in Sect. 3 below, from a very different 
perspective.

As a final consistency check, let us compute the $a\to 0$ limit
of the two-loop correlator \tolop. Using the asymptotics of the
Bessel function $I_0(x)\sim e^x/\sqrt{2 \pi x}$ for large $x$,
we get
\eqn\azerco{ \eqalign{
G(L_1,L_2\vert \tau,\sqrt{\Lambda},a\to 0)
&\sim {e^{-{1\over \tau a}(\sqrt{L_1}-
\sqrt{L_2})^2} \over \sqrt{4 \pi \tau a \sqrt{L_1L_2}}}\cr
&\to {1\over 2\sqrt{L_1}}\delta(\sqrt{L_1}-\sqrt{L_2})=\delta(L_1-L_2), \cr}}
where we have used the limit 
$\lim_{s\to 0^+} e^{-x^2/s}/\sqrt{\pi s}=\delta(x)$. We recover therefore 
the expected flat world-sheet result (also stated as 
$Z_{\ell_1,\ell_2}(T\vert 1,0)=\delta_{\ell_1,\ell_2}$).

\subsec{Other Boundary conditions}

In \AMB, where boundary conditions were chosen to be periodic, the result 
for the two-loop correlator in the scaling limit 
is very similar to \lapinv\ except that $a=1$ (no curvature) and more 
importantly that the modified Bessel function $I_0$ is replaced by $I_1$. 
We have a very simple explanation for this fact, that involves computing the 
$p$-seam loop correlator which we will define now. Up to now, we only 
considered triangulations with ``staircase'' boundary conditions.
Let us denote these triangulations as being of type (I).  Starting from
such a type (I) triangulation in its dual picture, 
a seemingly easy way to get periodic boundary 
conditions would be simply to identify the left and right staircase 
boundaries of the triangulation to construct a cylinder marked by a seam, 
remnant of the staircase. This construction is however problematic. 
Indeed, on each horizontal step of the left staircase boundary, we can have 
an arbitrary number of incoming lower half edges. Similarly, on any horizontal 
step of the right staircase boundary, we have an arbitrary number of incoming 
upper half-edges. When connecting these two staircases, there is no well 
defined natural prescription for deciding how to place these lower and upper 
half edges with respect to one-another. 
\fig{An example of triangulation of type (I) and a triangulation of type (II).
The boundary chaplets serve as seams to connect type (I) to type (II) 
triangulations. We give an example of cylinder made of two glued 
triangulations. Any marked point (black dot) on the lower (resp.\ upper) loop 
defines a unique chaplet of triangles pointing up (resp.\ 
pointing down).}{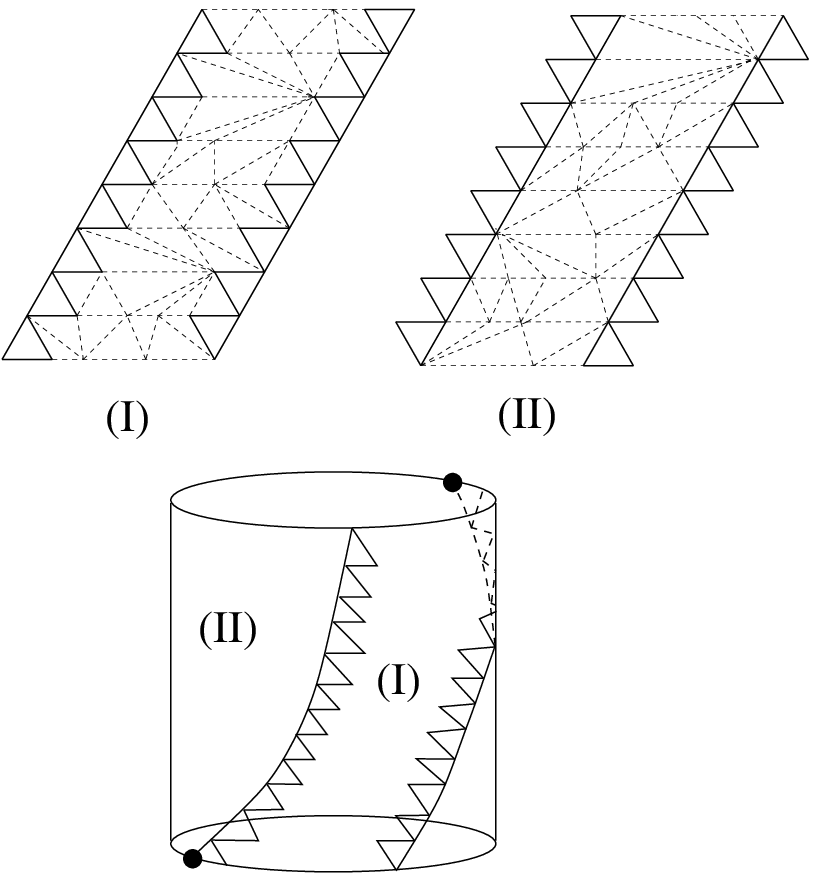}{8.cm}
\figlabel\seam
To overcome this problem , we need 
to introduce a new type of triangulations, which we will call type (II),
and  which can be glued without ambiguity to a type (I) triangulation.
Let us first see how the existence of a ``staircase'' in the dual 
representation translates into the original (i.e. non-dual) representation
made of triangles. Each leftmost lower half-edge of the dual picture
becomes a triangle pointing up. Each such leftmost triangle is attached by
its lower left vertex to the upper vertex of the leftmost triangle in the layer
just below, so as to form a chaplet (see Fig.\seam ). Similarly, the right 
staircase translates into a chaplet of triangles pointing down, each triangle
being attached by its upper right vertex to the lower vertex of the
triangle in the layer just above. The ``space'' between these two chaplets 
is filled with strips of triangles of arbitrary length, including 
strips of length zero  corresponding to the case where the two chaplets
would be in contact. The two chaplets will serve as seams in our 
construction. In order to glue our type (I) triangulation to a type
(II) triangulation, this type (II) triangulation must itself have boundaries 
made of chaplets but now with the reverse convention, i.e. with
a chaplet of triangles pointing up as its right boundary and a chaplet
of triangles pointing down as its left boundary (see Fig.\seam).
Again, the space between these chaplets is to be filled with
strips of arbitrary, possibly zero, length.
It is clear that we can glue a type (I) triangulation to a type
(II) triangulation on any side by simply superimposing the 
left (resp.\ right) chaplet of the former to the right (resp.\ left)
chaplet of the latter. We can thus construct a $p$-seam loop
correlator by gluing $p$ triangulations of alternating type (I) and (II), 
and, assuming that $p$ is even, gluing the last triangulation to 
the first one so as to form a cylinder (see Fig.\seam\ for
$p=2$). If the number $p$ of
triangulations is odd, we can build an open object with $p-1$ seams
and either type (I) or type (II) boundary conditions, depending on
whether to two extremal triangulations are of type (I) or of
type (II). This construction gives a nice a posteriori explanation for 
our choice of weight $\sqrt{g}$ per boundary half-edge: this is because 
the edges (or the triangles) have to be identified by pairs to  form 
the seam, hence the weight is right to get a factor $g$ per half-edge 
of the seam.
\fig{The one-to-one correspondence between type (I) and type (II)
triangulations. Starting from a type (I) triangulation (a), we move
each slice one step to the left with respect to the slice just below
to obtain (b). We then send the lower left and upper right vertices of
the boundary triangles thus released in each slice to respectively
upper left and lower right position
to obtain the type (II) triangulation (c), or equivalently
(d). The passage from (a) to (d) is clearly invertible.}{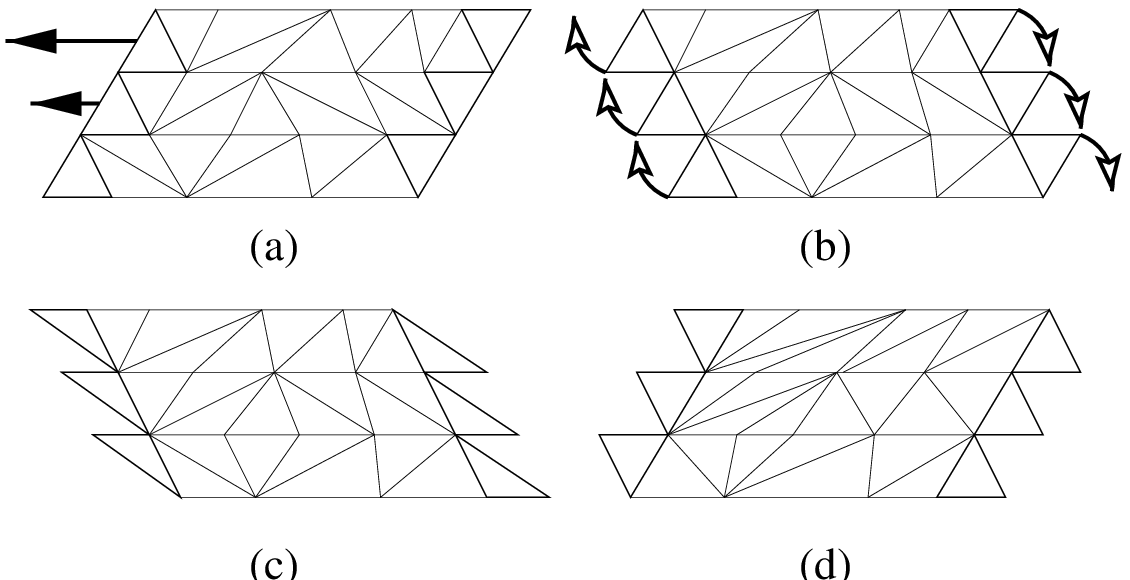}{8.cm}
\figlabel\ItoII
We have computed in Sect. 2.1 the transfer matrix for type (I)
triangulations, given by \weget . A remarkable result is that, although
type (II) triangulations are different from type (I) triangulations,
the transfer matrix for type (II) triangulations is {\it identical} 
to that of type (I), i.e. 
\eqn\esbctm{T^{(II)}_{i,j}(g,a)= T^{(I)}_{i,j}(g,a) = T_{i,j}(g,a),}
with $T_{i,j}(g,a)$ given by \weget .
To understand this property, we remark that starting from a type
(I) triangulation, we can deform it into a type (II) triangulation
by simply letting each time slice move one step to the left with respect
to the slice just below, and by sending the lower left and upper right
vertices of the boundary triangles so released in each slice, to
respectively upper left and lower right positions (see Fig.\ItoII). 
This transformation is clearly invertible, hence the announced result
\esbctm .

Thanks to the above construction, we may define and compute the 
$p$-seam two-loop correlator by juxtaposing side by side $p$ copies 
of alternating (I) and (II) triangulations, with a total lower 
(resp.\ upper) loop length of $\ell_1$ (resp.\ $\ell_2$), in which 
we simply have to identify the consecutive chaplet boundaries 
(again, the boundary weights produce the desired weight $g$ per 
half-edge of seam).
Assuming $p$ even and gluing the last triangulation to the first one, 
this leads to the $p$-seam two-loop correlator
\eqn\pseam{ Z^{(p)}_{\ell_1,\ell_2}(T\vert g,a)=
\big[ \big( T(g,a)\otimes T(g,a) \otimes ... \otimes
T(g,a) \big)^T\big]_{\ell_1,\ell_2} }
where there are $p$ factors in the tensor product, and the matrix element 
$(\ell_1,\ell_2)$ actually corresponds to the sum over all $p$-tuples of 
pairs of indices $(m_i,n_i)$ for each $T(g,a)$, with $\sum_i (n_i-1/2)=\ell_1$ 
and $\sum_i (m_i-1/2)=\ell_2$.
In terms of generating functions, the $p$-seam correlator \pseam\ is 
simply generated by: 
\eqn\pscore{\theta_{q,T}^{(p)}(x,y\vert \lambda)=\left(
{\theta_{q,T}(x,y\vert \lambda)\over \sqrt{xy}}\right)^p, \qquad p\ \hbox{even}
\ .}  
For odd $p$, gluing $p$ triangulations of alternating type (I)
and (II) leads to an open object of either type (I) or type (II)
with the two-loop correlator now generated by:
\eqn\pscoro{\theta_{q,T}^{(p)}(x,y\vert \lambda)=\sqrt{xy} \left(
{\theta_{q,T}(x,y\vert \lambda)\over \sqrt{xy}}\right)^p, \qquad p\ \hbox{odd}
\ .}  

Let us now perform the scaling transformations \scatran\ on \pscore\ or
\pscoro .We obtain the $p$-seam scaling function (note that it has required 
a multiplication by $\alpha^p$ to produce a finite limit)
\eqn\scat{ \Theta^{(p)}(X,Y\vert \tau,\sqrt{\Lambda},a)
=\Theta(X,Y\vert \tau,\sqrt{\Lambda},a)^p.}
We get the corresponding two-loop correlator by applying the inverse 
Laplace transform,
and performing integrations analogous to those of \lapinv:
\eqn\lapvin{\eqalign{
G^{(p)}&(L_1,L_2\vert \tau,\Lambda,a)=\int_{-i\infty}^{i\infty}dXdY 
e^{L_1X+L_2Y} \Theta(X,Y\vert \tau,\sqrt{\Lambda},a)^p \cr
&=\Lambda\int_{-i\infty}^{i\infty} du dv e^{\sqrt{\Lambda}(L_1u+L_2v)} 
\left({2\phi\over\sqrt{\Lambda}((u+1)(v+1)-\phi^2(u-1)(v-1))}\right)^p\cr
&=\int_{-i\infty}^{i\infty} {du \over (p-1)!} 
{(2\phi)^p\Lambda e^{\sqrt{\Lambda}L_1u}
\over (\sqrt{\Lambda}(u(1-\phi^2)+(1+\phi^2)))^p}
{d^{p-1}\over dv^{p-1}} e^{\sqrt{\Lambda} L_2 v} 
\bigg\vert_{v=-{u(1+\phi^2)+(1-\phi^2)\over u(1-\phi^2)+(1+\phi^2)}}\cr
&={\sqrt{\Lambda}(2\phi L_2)^p\over (p-1)! L_2}\int_{-i\infty}^{i\infty} du 
{e^{\sqrt{\Lambda}\left(L_1u-
L_2{u(1+\phi^2)+(1-\phi^2)\over u(1-\phi^2)+(1+\phi^2)}\right)} \over 
(u(1-\phi^2)+(1+\phi^2))^p} \cr
&={\sqrt{\Lambda}(2\phi L_2)^p\over (p-1)! L_2}
\sum_{k\geq 0} {1\over k!}
\bigg({4\phi^2\over 1-\phi^2}\sqrt{\Lambda}L_2\bigg)^k 
\int_{-i\infty}^{i\infty} du {e^{\sqrt{\Lambda}(L_1 u-L_2
{1+\phi^2\over 1-\phi^2})}
\over (u(1-\phi^2)+(1+\phi^2))^{p+k}}\cr
&={2\phi\sqrt{\Lambda}(\sqrt{L_1L_2})^{p-1}\over(1-\phi^2) (p-1)!}
e^{-\sqrt{\Lambda}(L_1+L_2){1+\phi^2\over 1-\phi^2}}
\sum_{k\geq 0}  
{\left( {2\phi\over 1-\phi^2} \sqrt{ \Lambda L_1L_2 } \right)^{2k+p-1}
\over k! (k+p-1)!}, \cr}}
for $p\geq 1$. In terms of the rescaled variables, the $p$-seam two-loop
correlator reads
\eqn\resop{\eqalign{G^{(p)}&(L_1,L_2\vert \tau,\Lambda,a)\cr
&= {\sqrt{\Lambda}
(\sqrt{L_1L_2})^{p-1}\over  {\rm sinh}(a\tau \sqrt{\Lambda})
(p-1)!} e^{-\sqrt{\Lambda}(L_1+L_2)
{\rm cotanh}(a\tau\sqrt{\Lambda})}
I_{p-1}\left({2\sqrt{\Lambda L_1L_2} \over 
{\rm sinh}(a\tau\sqrt{\Lambda})}\right), \cr}}
where $I_m$ denotes the $m$-th modified Bessel function, defined by the
series $I_m(x)=\sum_{k\geq 0} (x/2)^{m+2k}/(k! (m+k)!)$.

When $p=1$ this reduces to \tolop , as it should. More interestingly,
when $p=2$, in the pure gravity case $a=1$, the expression reduces, 
up to a factor of $L_2$, to the result for the two-loop correlator in periodic 
boundary conditions with one marked point on the lower loop \AL , i.e.
equivalently, the two-loop correlator in periodic boundary conditions 
with one marked point on the lower loop and one marked point
on the upper loop (note that marking a point on the external loops
simply amounts to multiplying by the length of this loop in the correlator).

On can easily explain this coincidence by noticing that, starting from a 
cylindric world-sheet with a marked point, say, on the lower loop, 
this marked point defines a unique chaplet of up triangles crossing the
cylinder from the lower to the upper loop. Indeed this point is
the lower left vertex of a unique up triangle in the first slice,
whose upper vertex is itself the lower left vertex of a unique triangle
in the second slice, and so on (see Fig.\seam ). Similarly, any marked point 
on the upper loop defines a unique chaplet of triangles pointing down.
Thus marking a point on the upper loop and one on the lower loop
amounts to marking two chaplets, one made of up triangles and one made
of down triangles. Such chaplets cannot intersect and thus divide
the cylinder into two triangulations, one of type (I) and one of 
type (II). This explains the connection between the two-seam 
two-loop correlator and the periodic loop correlator.
Note that this nice property breaks down as soon as $p > 2$ 
because, for more than one point on each external loop, we cannot
guaranty for arbitrary chosen marked points that the corresponding
chaplets alternate along the cylinder between ``up'' and ``down''
chaplets, which is crucial to keep our interpretation as $p$-seam
correlator. In particular two successive chaplets of the same 
(up or down) type can now merge into a single chaplet.

Note finally that we have the following relation between our $(2p+2)$-seam
correlator and the amplitude $A_p$ (with $p+1$ a type of winding number)
obtained by Nakayama in a continuum calculation using proper time
gauge \NA
\eqn\wind{G^{(2p+2)}(L_1,L_2\vert \tau,\Lambda, a=1)=
{(L_1 L_2)^p \sqrt{L_1 L_2}\over (2p+1)!}A_p.}

\newsec{Lorentzian Triangulations as Random Walks}

In this section, we will discuss the equivalence between Lorentzian
gravity configurations and ordinary random walk (RW) configurations
in one dimension. This connection will allow us to re-phrase the 
quantities computed above in the language of RW statistics. For instance 
we will see how the two-loop correlator of Lorentzian gravity relates to a 
well-known generating function for the large excursions of a Brownian motion. 
This equivalence will also explain why Lorentzian triangulations have fractal
dimension 2 and
why introducing a curvature weight 
(i.e. taking $a\ne 1$) cannot change the continuum large scale properties 
of Lorentzian gravity, as found in Sect.\ 2.4. Finally, the RW 
picture will also provide a very direct derivation of the crucial property 
\addi\ for the product of two transfer matrices. 

\subsec{Equivalence between Lorentzian triangulations and random walks}

The random triangulations describing Lorentzian gravity can be seen as 
deformations of a regular triangular lattice. More precisely, if we
view the regular triangular lattice as made of regular time slices of 
{\it alternating}
up and down triangles, then removing some of these triangles and gluing
together 
the remaining ones along their time-like edges within each strip leads
to a triangulation like that of Fig.\lorgra. Conversely, we can, 
starting from a random triangulation of this type, insert a number of 
additional triangles to make it regular. It is therefore possible to 
visualize a random triangulation as a regular one with two types of 
triangles: ``real'' triangles which are to be kept and form the triangulation
and ``virtual'' triangles which have to be shrunk or removed. 

This splitting procedure into real and virtual triangles can be made 
according to a well defined procedure which we shall now describe,
and which takes the 
form of a one-to-one correspondence between Lorentzian  triangulations 
and directed random walks drawn on the regular triangular lattice.

\fig{Starting from a directed random walk (a) drawn on the regular triangular 
lattice, we associate to each ascending step of the walk the pair of triangles 
lying just below it (grey triangles). We then eliminate all the other 
triangles and glue the grey triangles together 
along their time-like edges inside each strip,
thus forming a random Lorentzian triangulation (b). The $0$ and $T+1$ slices 
(light-grey triangles) are finally removed (c).}
{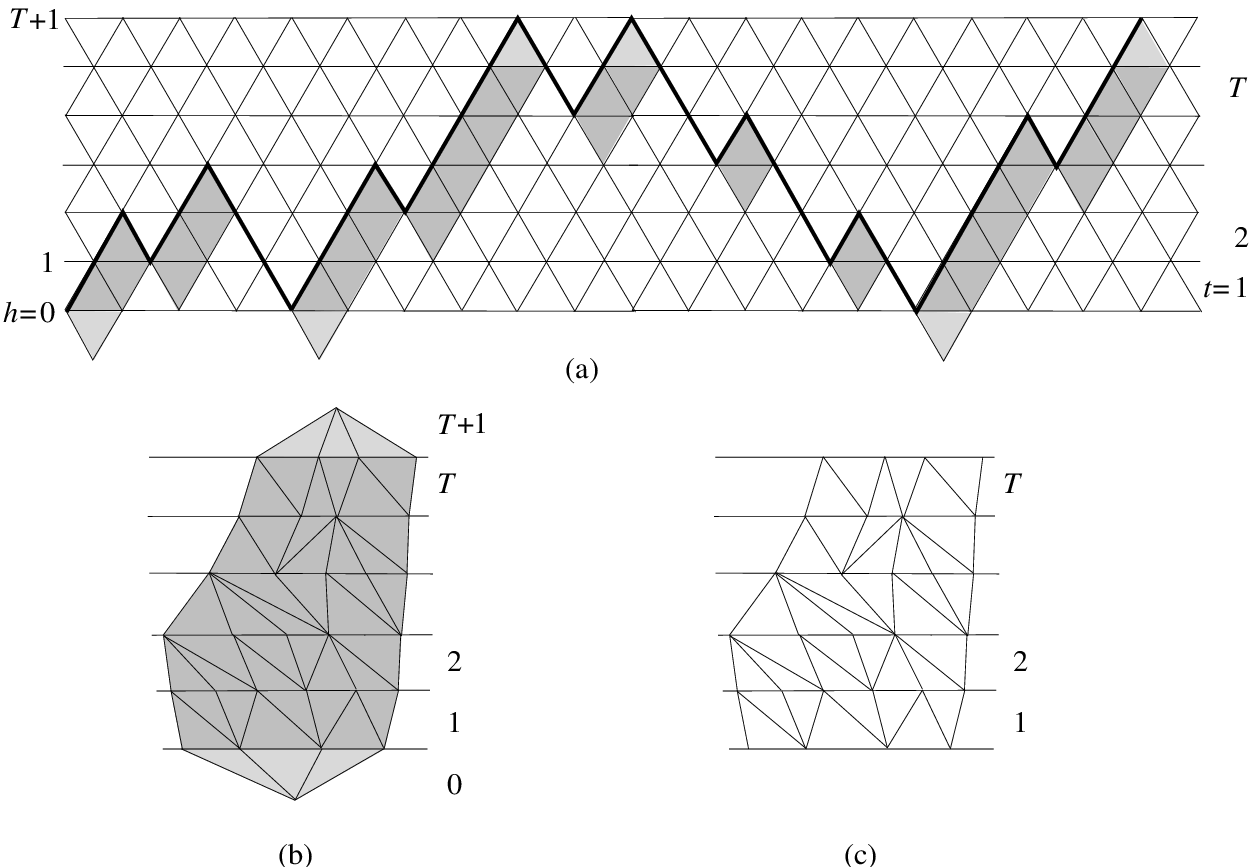}{12.cm}
\figlabel\RWtoLG
The equivalence is as follows:

Consider a directed random walk drawn on the regular lattice, starting at
``height'' $h=0$, making $\Delta h=\pm1$ steps to the right and ending at 
height $h=T+1$ after $S$ steps (see Fig.\RWtoLG). Since the walk is directed, 
the horizontal direction can be viewed as well as an effective time dimension 
for the random walk whose motion takes place in the one dimensional vertical
$h$-direction. This height variable $h$ will however correspond to the time 
variable $t$ of the triangulation, with each time slice $t$ lying between 
heights $h=t-1$ and $h=t$. The random walk is moreover required to stay 
confined within the strip $0\le h\le T+1$. Now to each elementary 
{\it ascending} step of the RW between heights $h$ and $h+1$, we associate 
the {\it pair of triangles} of the regular lattice made of the up triangle 
lying immediately to its right in the time slice $h+1$ and the down triangle 
lying just below in the slice $h$ (see Fig.\RWtoLG). These triangles will 
be the ``real'' triangles to be kept in the triangulation. 
Removing all the other triangles and gluing the real ones together 
along their time-like edges, we end up 
with a random Lorentzian triangulation of width $T$, and decorated by $l_1$ 
down triangles in time slice $t=0$ and $l_2$ up triangles in time slice $t=T+1$. 
These extremal time slices are then removed to recover the relevant
triangulation.
Note that from the above construction where we add triangles only to the 
right of the ascending steps, it follows that 
the resulting triangulation satisfies the staircase
condition of Sect.\ 2.1\foot{Note that the triangles can be viewed as well as
being added on the left side of the descending slopes of a directed walk, now 
going from height $h=T$ and reaching height $h=-1$ after $S$ steps to the 
left.}.

\fig{Starting from a Lorentzian triangulation satisfying staircase boundary
conditions (a), we go to the dual configuration (b) and attach each vertical
dual bond to the bond to its left in the preceding strip, as shown by the 
arrows,
to get a tree structure (c). Following the contour of the trees from the lower 
left branch to the upper right one (black points), the sequence of ascents and
descents can be translated into a directed random walk (d).}{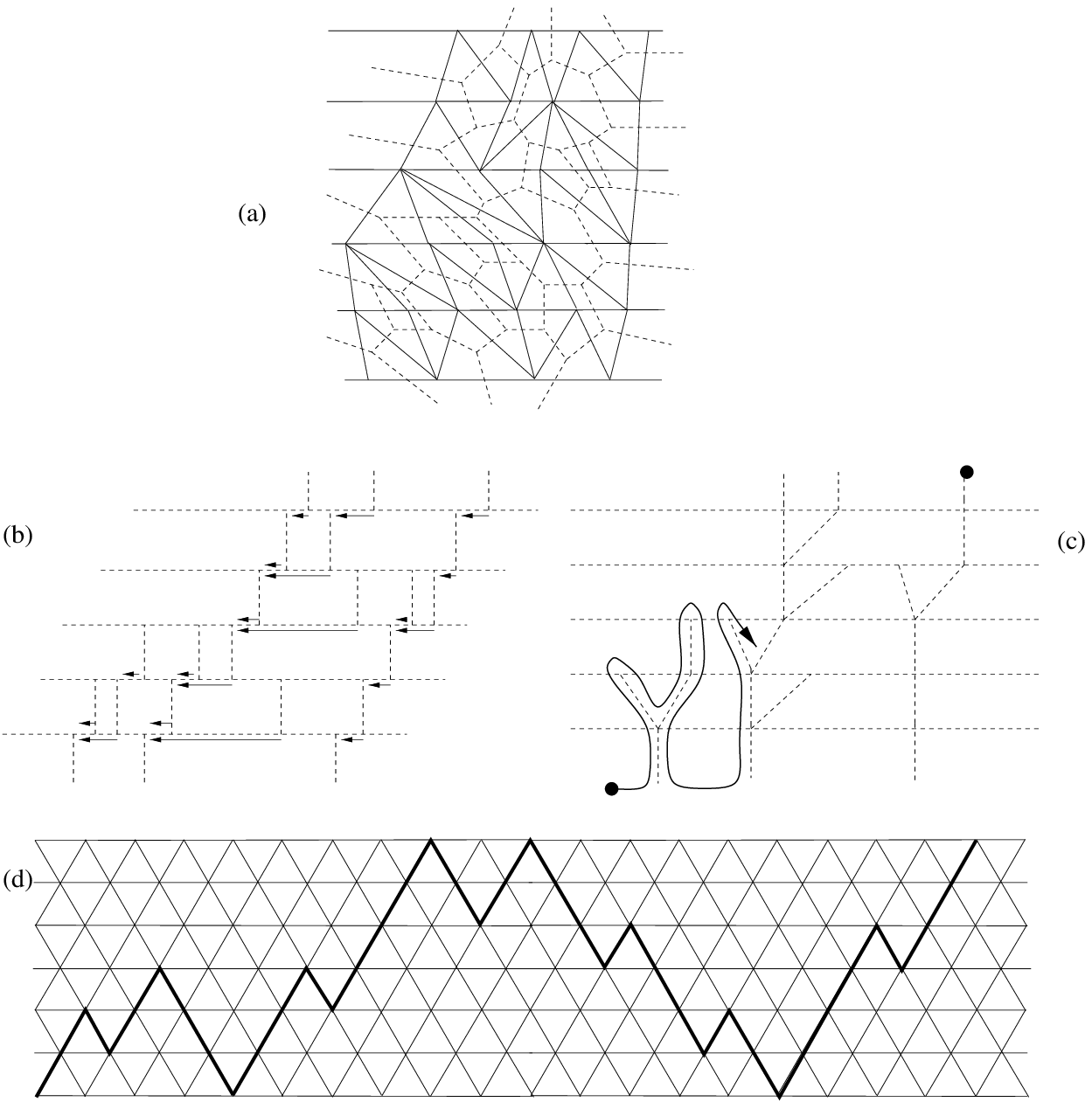}{12.cm}
\figlabel\LGtoRW

We can thus associate to each directed RW of the type described above a 
Lorentzian
triangulation with staircase boundary conditions. Let us now describe how,
starting now from such a Lorentzian triangulation, we recover the associated RW.
This reverse construction can be performed in three steps as shown in 
Fig.\LGtoRW . The first step is to go to the dual representation of the 
triangulation 
(Fig.\LGtoRW -b). In a second step, the dual world-sheet configuration can be 
transformed into a tree configuration (Fig.\LGtoRW -c) by attaching each 
vertical dual bond to the bond sitting in the strip just below and to its left. 
In a third 
step, this tree-like structure is transformed into a directed random walk 
(Fig.\LGtoRW -d) by simply following the contour of the trees from the lower 
left branch to the upper right one and making a step $\Delta h=+1$ for each 
ascent along a branch and a step $\Delta h=-1$ for each descent along a branch. 
It is easy to check that the resulting RW is precisely the expected one.

Our main result here is thus a {\it one-to-one correspondence} between
directed random walks and Lorentzian triangulations, which in turn implies
a direct equivalence for the various quantities describing them. Some of these 
relations will be described in the next section. Let us simply note here that 
we can deal with all types of boundary conditions. The staircase boundary
condition described in this section corresponds to random walks starting at 
$h=0$ 
and ending at height $h=T+1$. The case of walks starting and ending at $h=0$
would describe periodic boundary conditions with a marked point in the lower 
loop of the triangulation.

\subsec{Two-loop correlator as a generating function for large excursions}

Let us now re-interpret some of the natural statistical properties
of random Lorentzian triangulations in the language of random walks.

Since we associate a pair of triangles to each ascending step, 
the number $N_t$ of triangles is directly related to the length
$S$ of the walk by:
\eqn\nule{N_t+(l_1+l_2)=S+T+1.} 
Here $N_t$ counts all the triangles with $1\le t\le T$, ignoring
the $(l_1+l_2)$ triangles in the slices $t=0$ and $t=T+1$ which are
removed in the construction of the triangulation. We also used
the fact that there are $T+1$ more ascending steps than descending
steps in the walk.
Setting $a=1$ for the time-being, the triangulation comes
with a factor $g^{N_t-T}$ due to the particular treatment of
the extremal left and right triangles in each of the $T$ slices.
The corresponding weight $g^{S-(l_1+l_2-1)}$ can be interpreted as 
a fugacity $g$ {\it per step} for the random walk, together
with a reflection factor $(1/g)$ for each of the $l_1$ contacts at $h=0$
or the $l_2$ contacts at $h=T+1$, except for the last contact. Note that for 
the critical value $g=1/2$, we can view these weights as a 
probability $1/2$ for the walk to make a step up or down inside 
the strip and a probability $1$ for the walk to make an up (resp.\ 
a down) step at the $h=0$ (resp.\ $h=T+1$) boundary, i.e. we have 
an unbiased random walk between two reflecting walls.
 
The sizes $l_1$ and $l_2$ of the bottom and top loops
of the triangulation correspond precisely to the number of
contacts at $h=0$ and at $h=T+1$ respectively.
We can thus reinterpret the two-loop correlator 
$Z_{l_1,l_2}(T\vert g,a=1)$ as the generating function
\eqn\proba{Z_{l_1,l_2}(T\vert g,a=1)={1\over (2g)^{l_1+l_2-1}}
\sum_{S\geq 0} (2g)^S P_T(l_1,l_2,S)}
where $P_T(l_1,l_2,S)$ is the probability for an unbiased
walker starting at $h=0$ and evolving between two reflecting
walls at $h=0$ and $h=T+1$ to reach the wall $h=T+1$ in $S$
steps after exactly $l_1$ contacts (including the first one) at the
bottom wall and $l_2$ (including the last one) at the top wall.

More generally, if we re-instate the factor $x$ (resp.\ $y$) per 
space like link in the bottom (resp.\ top) loop, the triangulation
generating function $\theta_T(x,y\vert g,a=1)$ (given by
Eq. \tetn\ for $a=1$, i.e. $q=(1-\sqrt{1-4g^2})/(2g)$ and
$\lambda=q^2$) becomes the
generating function for random walks between two walls with a weight $g$ per 
step and a weight $x/g$ (resp.\ $y/g$) per contact to the bottom 
(resp.\ the top) wall. In the RW language, we call each portion of walk 
between two {\it successive} contacts an ``excursion''. Since the walk 
is a Markov process, each of these excursions is an independent random 
object. Excursions here are of four types: bottom-bottom and top-top 
excursions, which have the same generating function (up to the exchange 
of $x$ and $y$), and  bottom-top and top-bottom excursions which also
have the same generating function. For instance the bottom-bottom 
excursion counts the walks starting and ending at $h=0$ and {\it staying 
below} $h=T$. The distribution probability for, say, the height and the 
length of bottom-bottom excursions is a basic quantity of the random walk
statistics and enters many different physical problems. Its continuous,
Brownian motion limiting scaling expression is well known, as are  
many other more involved excursion properties appearing in many different 
areas of physics. As an example, we would like to show here how the 
two-loop correlator of Lorentzian gravity directly relates to a well known 
probability distribution describing the diffusion of a particle in a 
one-dimensional random energy landscape. 
\fig{(a) A random walk entering the definition of $Z_{m+1,1}(T\vert g,a)$,
i.e. with $m+1$ contacts (here $m=4$) at $h=0$ and $1$ contact at $h=T+1$.
(b) The corresponding energy landscape obtained by erasing in (a) all the
ascending steps following immediately a contact at $h=0$. The deepest
minimum reachable by a diffusing particle (black dot) passing energy 
barriers of height strictly less than $T$ is at depth $m$ 
(white circle).}{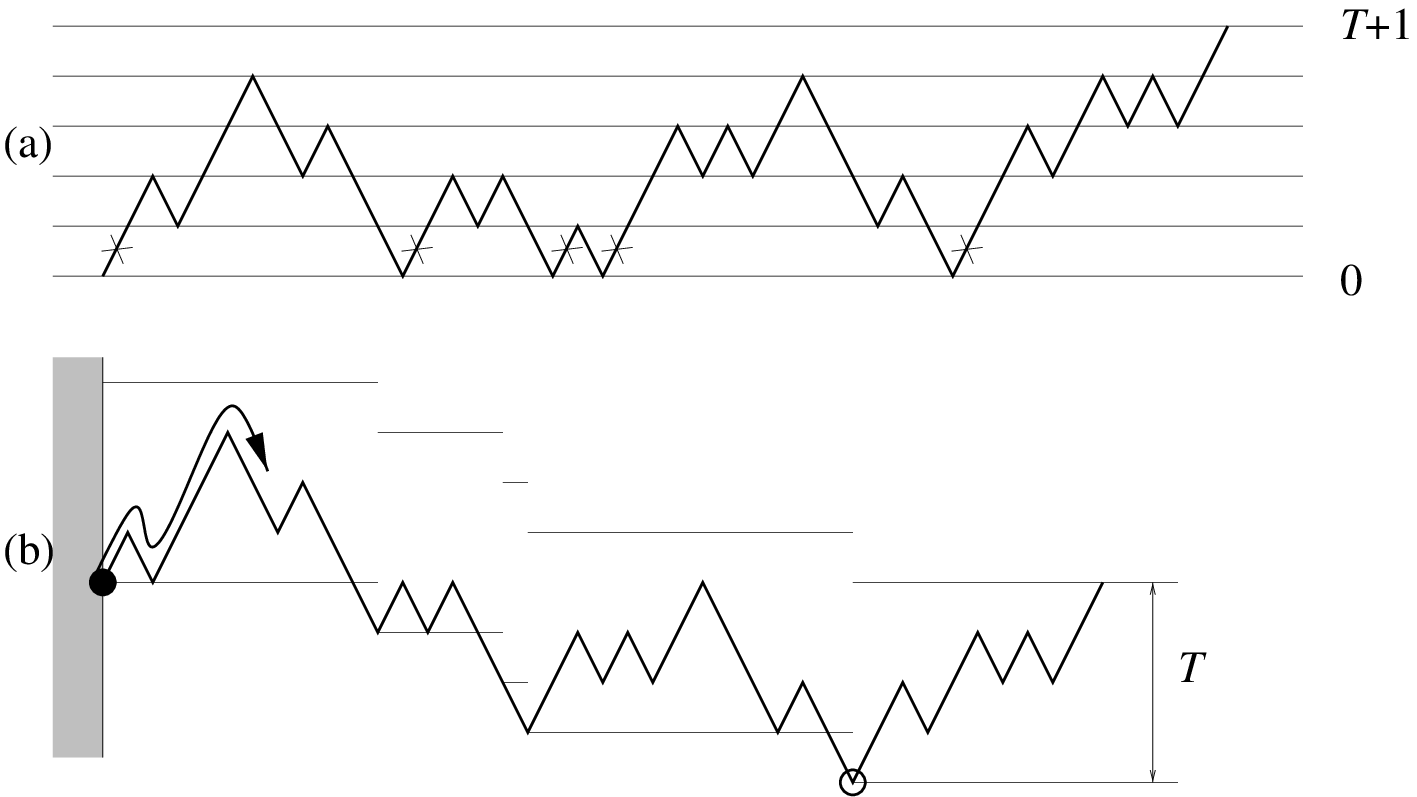}{10.cm}
\figlabel\sinaif
The diffusion of a particle in a random energy landscape can be modelled
by the so-called Sinai model \sinai, where a particle jumps to one of its 
two neighbouring
sites with a probability $p\propto {\rm exp}(-\Delta U)$ involving the energy
difference $\Delta U$ with this neighbouring site. If we represent
the energy landscape as a random walk, with the height $h\propto U$ 
corresponding to the energy, a bottom-bottom excursion describes an energy
barrier of height less than $T$. Exact expressions for the large
time properties of the diffusion have been obtained and it was recognized 
that a good way to recover these expressions is to assume that the particle 
at time $t$ is {\it localized} precisely in the lowest energy minimum it 
can reach by passing all the energy barriers of height less 
than a certain maximal size $T(t)$, with the relation 
$T(t)\propto \ln(t)$. Let us for instance consider the example 
of  Fig.\sinaif\ (b) where a particle, starting at a wall, jumps
to the right  by passing all barriers of height strictly less than $T$, looking
for lower and lower minima to the right until it reaches a barrier
of height larger than
or equal to $T$. How far and how deep does the particle go? From 
\proba , we know that the quantity 
\eqn\sinai{Z_{l_1=m+1,l_2=1}(T\vert g,a=1)={1\over (2g)^{m+1}}
\sum_{S\geq 0} (2g)^S P_T(m+1,1,S)\ ,}
relates to the probability $P_T(m+1,1,S)$ that the walk reaches 
$h=T+1$ for the first time after $S$ steps and passes $m$ times at 
$h=0$. If we deform the walk by suppressing the first ascending step 
after each contact point at $h=0$, the quantity $P_T(m+1,1,S)$
becomes the  probability for the diffusing particle to have reached a depth
$m$  by passing barriers of length strictly less than $T$ and being 
blocked after $S-m-1$ steps by a barrier of height larger or equal
to $T$. Going to the continuous limit and using \tolop , we obtain
that the two-loop correlator 
\eqn\tolopsin{G(L,0\vert \tau=1,\sqrt{\Lambda},a=1)
= {\sqrt{\Lambda} \over {\rm sinh}(\sqrt{\Lambda})} 
e^{-\sqrt{\Lambda}L {\rm cotanh}(\sqrt{\Lambda})},}
should be equal to the Laplace transform 
\eqn\lapl{\hat{P}(L,\Lambda)\equiv \int_0^\infty dx P(L,x) 
e^{-\Lambda x},}
of the probability density $P(L,x)$ for the particle to reach the 
deepest minimum at depth $L$ and being blocked at a distance $x$ from
the wall, in the scaling limit of large $T$, $m$, and $S$ with $m/T=L$ 
and $S/T^2=x$ fixed\foot{Strictly speaking, the particle is blocked after
$S-m-1$ steps, not $S$, but this correction scales less rapidly 
than $T^2$ and is thus negligible.}. This probability is the
convolution of the probability $P_1(L,x_1)$ that the particle
reaches the deepest minimum at depth $L$ and distance $x_1$ from
the wall, and the probability $P_2(x_2)$ that the last excursion
to the blocking point is of length $x_2$. The expression \tolopsin\ 
is then a direct consequence of the following expressions
\eqn\sinexp{\eqalign{&
\hat{P}_1(L,\Lambda)\equiv \int_0^\infty dx_1 P_1(L,x_1)
e^{-\Lambda x_1}=e^{-\sqrt{\Lambda}L {\rm cotanh}(\sqrt{\Lambda}),}\cr
&\hat{P}_2(\Lambda)\equiv \int_0^\infty dx_2 P_2(x_2)
e^{-\Lambda x_2}={\sqrt{\Lambda}\over {\rm sinh}(\sqrt{\Lambda})}\cr},}
which are well known in the context of the Sinai model \CG.

The RW equivalence also provides an explanation for
the fractal dimension 2 of Lorentzian triangulations.
It is well-known that the Brownian motion has fractal 
dimension 2, meaning that the time (represented here 
by the length $S$ of the walk) needed to travel over 
a distance $T$ (here the height $h\sim T$) scales as $T^2$.
Since $S$ corresponds in our equivalence to the number $N_t$
of triangles, i.e to the area $A$ of the triangulation, we 
obtain that $A\sim T^2$ and thus get a fractal dimension 2 
for Lorentzian triangulations. Note that the fact that
$l_1$ (or $l_2$) scales as $T$ is also clear in the Sinai 
model language (with $l_1\sim m$) where it means that the depth 
reached by passing maximal barriers of order $T$ is itself of 
order $T$.

\subsec{Curvature and universality} 

Let us now re-instate a curvature weight $a\ne 1$ in the problem,
and let us see how it translates into the random walk language. 
As for the equivalence with tessellations of triangles and squares
of Sect. 2.4, we can regroup the up and down triangles
in each strip as follows. One first re-groups all pairs
of triangles made of an up triangle immediately followed to 
its right by a down triangle and assign a weight $g^2$ to these pairs.
The remaining up triangles necessarily have an up triangle to
their right. We thus assign them a weight $ga$ accounting for 
the fugacity $g$ per triangle and the curvature weight $a$ for
their interaction with their right neighbour. Similarly, each 
remaining down triangle necessarily has a down triangle to its left
and again receives a weight $ga$ for the same reasons. With this 
construction, all the weights have been collected properly and
we simply need an extra factor $1/g$ to correct for the special boundary
weights. Otherwise stated, we can assign a weight $ga$ for each 
triangle and correct with a factor $1/a^2$ for each pair made of an up 
triangle followed by a down one and a global $1/g$ factor. 
This way of correcting is alternative to that of Sect. 2.4
consisting in adding squares in the game.
Denoting by $P$ the total number of up-down pairs in the whole
triangulation, we get a weight:
\eqn\facrt{{1\over g^T}{\left(ga\right)^{N_t}\over 
\left(a^2\right)^P}\ .}
Now in the RW picture, it is easy to see that up triangles lie 
immediately to the right of an ascending step, and that they are 
followed by a down triangle if and only if the next step is also 
ascending. In other words, pairs of up-down triangles correspond 
exactly to pairs made of two successive ascending steps. 
Denoting by $n_{++}$ (resp.\ $n_{+-}$, $n_{-+}$, $n_{--}$) the
pairs of successive ascending-ascending (resp.\ ascending-descending,
descending-ascending, descending-descending) steps, we have
the relations
\eqn\reln{\eqalign{
n_{+-}&=n_{-+},\cr
P&=n_{++},\cr 
T&= n_{++}-n_{--},\cr
S&= n_{++}+n_{+-}+n_{-+}+n_{--}\ .\cr}}
The first relation simply states that there are as many summits
as valleys in the RW landscape. Thanks to these relations
and to Eq. \nule , we can rewrite \facrt\ as 
\eqn\faccur{{g^{S-(l_1+l_2-1)} a^{n_{++}+n_{+-}+n_{-+}+n_{--}+T-
(l_1+l_2-1)}\over a^{n_{++}+n_{--}+T}}
=g^S {1\over \left(ga\right)^{l_1+l_2-1}} a^{n_{+-}+n_{-+}}\ .}
The introduction of $a\ne 1$ has thus two effects
in the RW weight: (i) it changes the interaction of contact with the walls
from $(1/g)$ to $(1/ga)$ and (ii) introduces a factor $a$ for
each {\it change of slope} in the walk. Note that the two
effects cancel exactly at the walls so that the net contribution
is simply a factor $a$ for each change of slope {\it inside}
the strip. Such a weight is nothing but an {\it extrinsic} 
curvature energy $E\propto - \ln(a)$ for the one dimensional random walk.
It is thus remarkable that the {\it intrinsic} curvature weight of
two-dimensional Lorentzian gravity translates into an {\it
extrinsic} one-dimensional curvature weight for the equivalent random walk
problem. Note that according to whether $a$ is less or larger than one,
one has a positive (favoring straight lines) or negative
(favoring U-turns) curvature elastic constant. 

It is well known that extrinsic curvature is an irrelevant 
perturbation for the large scale properties of the random 
walk. At most it introduces a finite correlation length below
which the random walk is rigid (positive curvature)
or crumpled (negative curvature) but at scales larger than
this correlation length, the curvature plays no role. Up
to an appropriate rescaling, one thus recovers asymptotically
the same continuum limit as without curvature energy. This is 
precisely what we observed in \tolop. 

\subsec{Product of two transfer matrices}

{}From the RW picture of Lorentzian triangulations, we can easily 
re-derive the relation \addi\ for the product of two transfer 
matrices. Indeed, a two-matrix product element like
$[T(g,a)T(g',a')]_{i,j}$ corresponds to a sum with 
appropriate weights over all Lorentzian triangulations made of 
exactly {\it two} time slices, with a fixed number $i$ of up 
triangles in their lower slice, a fixed number $j$ of down triangles 
in their upper slice, and with staircase boundary conditions. 
In the RW representation, we thus have to consider all the random
walks in a strip of height three $0\le h\le 3$, starting at $h=0$,
ending at $h=3$ and having exactly $i$ contacts at $h=0$ and
$j$ contacts at $h=3$. We thus have to sum over all these RW with
a fixed number of contacts at the top and at the bottom of
the strip $0\le h\le 3$. 
\fig{The decomposition of a walk of height $T+1=3$ into blocks. Each block 
is a portion of the walk between two successive contacts at the walls
$h=$ or $h=3$. The blocs are of four types according to the 
four possibilities (1) bottom-bottom, (2) top-top, (3) bottom-top and
(4) top-bottom for the two contacts at their extremities.}{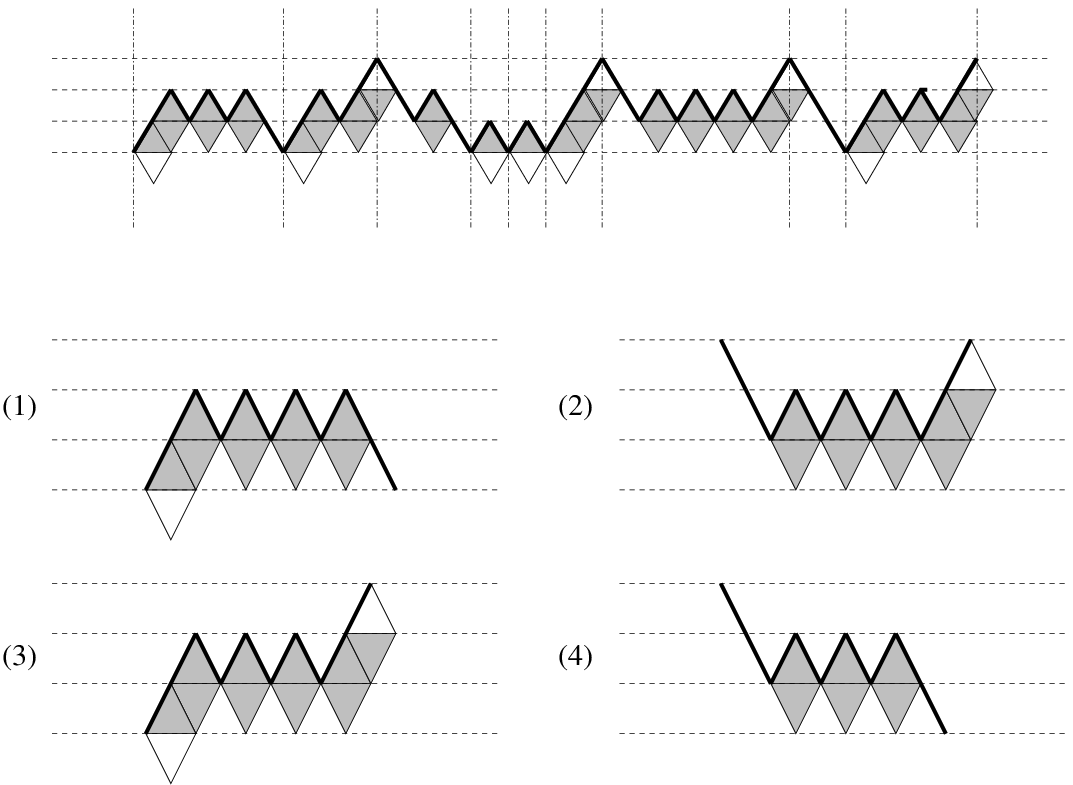}{10.cm}
\figlabel\blocs
Now we can decompose any such random walk
into $i+j-1$ ``blocks'' corresponding to the $i+j-1$ portions of the 
walk between two successive contacts (see Fig.\blocs ). The blocks are
of  four types, numbered (1) to (4), according to whether these contacts are at
the bottom or the top of the strip, namely (see
Fig.\blocs ):  (1) bottom-bottom,  (2) top-top,  (3) bottom-top,  and (4)
top-bottom. These blocks
differ only by the (up or down) nature of their first and  last steps while the
intermediate steps form a saw-tooth sequence  $\Delta
h=\sigma,-\sigma,\sigma,-\sigma,\sigma,-\sigma,\cdots$, with $\sigma=1$ or $-1$, 
of arbitrary  (possibly zero) length $k$. The desired sum over the random
walks can be achieved by first choosing one of the $i+j-2 \choose i-1$ allowed
sequences  for the bottom and top contacts, then summing over all the random
walks  having this particular sequence of contacts, and finally summing over
all sequences. Alternatively, we can attach directly a global weight
individually to each of the four block types by summing over all possible
intermediate configurations, i.e. over all values of the lengths $k$.

If we are interested in the matrix-product $T(g,a)T(g',a')$, we
must attach a weight $ga$ to each ascending step in
the slice $0\le h\le 1$, a weight $g'a'$ to each ascending step
in the slice $2\le h\le 3$ and a weight $(gg'aa')$ to each ascending
step in the slice $1\le h\le 2$. We then must correct this weight
by a curvature factor $1/a^2$ for each sequence $h=0,1,2$ and a 
factor $1/a'^2$ for each sequence $h=1,2,3$. Gathering these
factors and summing over $k$, we obtain the block weights:
\eqn\blocwtwo{\eqalign{
w_2^{(1)}& = ga+{ga\over a^2}\sum_{k=1}^\infty (gg'aa')^k = {ga
+g^2(1-a^2)g'a'\over 1-gg'aa'},\cr
w_2^{(2)}& = g'a'+{g'a'\over a'^2}\sum_{k=1}^\infty (gg'aa')^k ={g'a'
+g'^2(1-a'^2)ga\over 1-gg'aa'},\cr
w_2^{(3)}& = {(ga)(g'a')\over a^2a'^2}\sum_{k=1}^\infty (gg'aa')^k 
= {g^2g'^2\over 1-gg'aa'},\cr
w_2^{(4)}& = \sum_{k=0}^\infty (gg'aa')^k = {1\over 1-gg'aa'}. \cr
}}
Note that {\it all} the weights have been gathered {\it inside}
the blocks and that consecutive blocks do not interact.
In order to have the commutation $T(g,a)T(g',a')=T(g',a')T(g,a)$,
it is sufficient that the four weights above are invariant
under the change $(g,a)\leftrightarrow (g',a')$. This is
readily the case for the weights $w_2^{(3)}$ and $w_2^{(4)}$, while
$w_2^{(1)}$ and $w_2^{(2)}$ are exchanged. The commutation condition
reads therefore:
\eqn\comrel{w_2^{(1)}=w_2^{(2)} \leftrightarrow {1-g^2(1-a^2)\over ga}
={1-g'^2(1-a'^2)\over g'a'}\equiv q+{1\over q},}
defining for the two couples $(g,a)$ and $(g',a')$ a common value 
of $q$ as in \quadrat .

To go further and show \addi , we must compare the product
$T(g,a)T(g',a')$ to a single matrix say $T(g'',a'')$. Let us
note that the same block decomposition as before can be achieved for random
walks in a strip of height two, describing one-slice triangulations
contributing to the elements of a single matrix $T(g'',a'')$. In this
case, the contact sequence fully specifies  the random walk and the four block
types correspond the the four  possible two step sequences:
(1) $\Delta h= +1,-1$, (2) $\Delta h=-1,+1$, (3) $\Delta h=+1,+1$
and (4)$\Delta h=-1,-1$. 
This leads immediately to the following block weights:
\eqn\blocwone{\eqalign{
w_1^{(1)}& = w_1^{(2)}=g''a'',\cr
w_1^{(3)}& = g''^2,\cr 
w_1^{(4)}& = 1.\cr
}}
Let us now consider the {\it same} sequence of blocks with $n^{(i)}$
blocks of type $(i)$, in both the product $T(g,a)T(g',a')$ and $T(g'',a'')$.
On one hand, the weight
contribution to $T(g,a)T(g',a')$ is: \eqn\totwtwo{{1\over
gg'}\left(w_2^{(1)}\right)^{n^{(1)}} \left(w_2^{(2)}\right)^{n^{(2)}}
\left(w_2^{(3)}\right)^{n^{(3)}}\left(w_2^{(4)}\right)^{n^{(4)}} \ ,}
where the pre-factor $1/gg'$ comes from the special weight of the
boundary triangles. On the other hand, 
the contribution to $T(g'',a'')$ is simply:
\eqn\totwone{{1\over g''}\left(w_1^{(1)}\right)^{n^{(1)}+n^{(2)}}
\left(w_1^{(3)} \right)^{n^{(3)}} \ .}
Using the commutation relation \comrel\ and the fact that 
$n^{(3)}=n^{(4)}+1$, i.e. the number of ``ascending" blocks is one
unit more than the number of ``descending" blocks for the random
walks that we consider, we can rewrite \totwtwo\ as:
\eqn\totsim{{1\over gg'w_2^{(4)}}\left(w_2^{(1)}\right)^{n^{(1)}+n^{(2)}}
\left(w_2^{(3)}w_2^{(4)}\right)^{n^{(3)}} \ .}
We get therefore obtain the product relation:
\eqn\comu{T(g,a)T(g',a')={g''\over gg'w_2^{(4)}}T(g'',a''),}
provided that (i) the commutation relation \comrel\ is satisfied and (ii)
we have the relation:
\eqn\relcom{\matrix{w_2^{(1)}=w_1^{(1)}& & g''={\displaystyle{gg'}\over 
\displaystyle{1-gg'aa'}} \cr 
& \leftrightarrow & \cr
w_2^{(3)}w_2^{(4)}=w_1^{(3)} & & a''={\displaystyle{ga+g'a'-gg'aa'
\left(q+{1\over q}\right)} \over \displaystyle{gg'}}& \cr}}
with $q$ as in \comrel\ above. It is straightforward to check that
the above values of $g''$ and $a''$ obey the relation:
\eqn\GAq{{1-g''^2(1-a''^2)\over g''a''}= q+{1\over q},}
and thus the three couples $(g,a)$,$(g',a')$ and $(g'',a'')$ correspond
to the same value of $q$. If we introduce the parameter $\lambda$
as in \poval\ for $(g,a)$ and $\lambda'$ for $(g',a')$, the value
of $\lambda''$ for $(g'',a'')$ is found to be $\lambda''=\lambda\lambda'$,
while the pre-factor in \comu\ disappears since $g''/ gg'w_2^{(4)}=1$.  
We have therefore proven the relation \addi\ in the language of
random walks. One should note that the power of this method comes
from the fact that we used only four basic objects, the four blocks,
each of which corresponds to a particularly simple family of walks, 
while the arbitrary sequence of the blocks plays no role. We expect
that we can take advantage of such a block decomposition to obtain
commutation relations for the transfer matrices of more involved 
problems of Lorentzian gravity, in particular including matter
degrees of freedom.

\newsec{Other Models}

\subsec{General construction: symmetric case}

Let us reconsider the expression \eveceval\ for the curvature transfer matrix
$T_q(\lambda)$ in terms of the generating functions $F_m$ of its eigenvectors. 
If we now start from some arbitrary orthonormal family $v^{(m)}$, $m=1,2,...$
of vectors generating some space $E$, and a given non-zero function
$f(\lambda)$, we may easily construct a commuting set of symmetric transfer
matrices $T(\lambda)$ with entries  \eqn\gegen{ T_{i,j}(\lambda)=
f(\lambda)\sum_{m\geq 1} v_i^{(m)} \lambda^{m-1} v_j^{(m)}, }
that satisfy the multiplicativity property
\eqn\mulgene{T(\lambda)T(\lambda')={f(\lambda)f(\lambda')
\over f(\lambda\lambda')} T(\lambda\lambda').}

We also remark that, given any set of orthonormal vectors, generated say by
$F_m(x)$, then the other set generated by $G_m(x)=F_m(x) \varphi(x)$ is also
orthonormal, provided we take
\eqn\phix{\varphi(x)=\prod_{j=1}^k \varphi(x\vert q_j)^{k_j} \ \ , \qquad
\varphi(x\vert q)={q-x\over 1-qx},}
in which $q_1,...,q_k$ are arbitrary complex numbers inside the unit disc and
$k_j$ some given positive integers. Indeed, the functions $\varphi(x\vert q)$
satisfy the inversion relation $\varphi(x\vert q)\varphi(1/x\vert q)=1$,
hence the contour integrals $\oint G_mG_{m+p}$ are computed exactly like
those for $F_m$, and the norm is unchanged. 
Note in particular that for $q=0$, $\varphi(x\vert
q)\propto x$ allows us to multiply $F_m(x)$ by any power of $x$.
This gives us a lot of freedom
in picking the model.

Let us now present a simple example, with the same vector space
$E=\IR\otimes\IR\otimes...$ we have considered so far.

\subsec{Integrable models of discrete 2D Lorentzian gravity with polygonal
tiles}

Let $q_1,q_2,...,q_k$ be generic distinct real numbers with $|q_i|<1$.
Let us introduce the function
\eqn\fuin{ \phi_m(x\vert q_1,...,q_k)=x \prod_{j=1}^k 
{(q_j-x)^{m-1} \over (1-q_j x)^{m} },}
for $m=1,2,...$
We claim that this generates the components of a family of orthogonal vectors
$w^{(m)}$ spanning $E$. The spanning property is clear by noticing that
a triangular change of basis maps the canonical one $K_m(x)=x^m$ onto
the $\phi_m$'s\foot{This is true because we assumed the $q$'s to be generic,
i.e. not related by any algebraic relation. It is easy to see 
for instance in the case
$k=2$, $q_1=q$, $q_2=-q$ which is non-generic, that $\phi_m$ is an odd
function of $x$, in which case only ``half'' of $E$ is generated.}.
To see why they are orthogonal, we simply have to compute the following
contour integral over the unit disc
\eqn\proint{
\oint {dx \over 2i\pi x} \phi_m(x\vert q_1,...,q_k)
\phi_{m+p}({1\over x}\vert q_1,...,q_k)= (-1)^k \oint {dx \over 2i \pi} x^{k-1}
\prod_{j=1}^k {(1-q_j x)^{p-1} \over (q_j-x)^{p+1} },}
For $p>0$, the integrand has no poles outside the unit disc, hence the
integral vanishes, and the corresponding vectors are orthogonal. 
When $p=0$, we find the normalizations 
\eqn\norma{\eqalign{ N(q_1,...,q_k)^{-2}&= \oint {dx \over 2i \pi} x^{k-1}
\prod_{j=1}^k {1\over (1-q_j x)(x-q_j) }\cr
&=\sum_{j=1}^k  {q_j^{k-1} \over 1-q_j^2} \prod_{i\neq j} {1\over
(1-q_iq_j)(q_j-q_i)},\cr}} 
that reproduce the one-$q$ result
$N(q)=\sqrt{1-q^2}$, and give for $k=2,3$
\eqn\othnor{\eqalign{
N(q_1,q_2)&=\sqrt{ (1-q_1^2)(1-q_2^2)(1-q_1q_2)\over 1+q_1q_2},\cr
N(q_1,q_2,q_3)&=\sqrt{
(1-q_1^2)(1-q_2^2)(1-q_3^2)(1-q_1q_2)(1-q_1q_3)(1-q_2q_3)
\over 1+q_1q_2+q_1q_3+q_2q_3- q_1q_2q_3(q_1+q_2+q_3)-(q_1q_2q_3)^2 }. \cr}}
The normalizations \norma\ allow us to
define orthonormal vectors $v^{(m)}=N(q_1,...,q_k)w^{(m)}$, generated
by $F_m(x\vert q_1,...,q_k)\equiv N(q_1,...,q_k)\phi(x\vert q_1,...,q_k)$.
This produces an interesting candidate for the transfer matrix of Lorentzian
gravity including triangles, squares,...,up to $2k+2$-gons, with the same fixed
staircase boundary conditions as before: 
\eqn\intcan{
\theta_{q_1,...,q_k}(x,y\vert \lambda)={f_{q_1,...,q_k}(\lambda) x y
N^2(q_1,...,q_k) \over \prod_{i=1}^k (1-q_ix)(1-q_iy) -\lambda \prod_{i=1}^k 
(q_i-x)(q_i-y)}. }
To obtain a well-normalized generating function, we must take $f$ of the form
\eqn\takef{f_{q_1,...,q_k}(\lambda)={g(1-\lambda(q_1...q_k)^2)\over
N^2(q_1,...,q_k)}.}
where $\sqrt{g}$ is the Boltzmann weight per boundary triangle, a free
parameter of the theory, and the generating function for the transfer matrix
elements finally reads
\eqn\tmkq{
\theta_{q_1,...,q_k}(x,y\vert \lambda)={g x y \over
\sum_{i,j=0}^k w_{i,j}(q_1,...,q_k\vert \lambda) x^i y^j}, }
where
\eqn\bolw{ w_{i,j}(q_1,...,q_k\vert \lambda)=(-1)^{i+j}{\sigma_i \sigma_j
-\lambda \sigma_{k-i} \sigma_{k-j}\over \sigma_0^2 - \lambda \sigma_k^2},}
where the $\sigma_j\equiv \sigma_j(q_1,...,q_k)$ are the symmetric
functions of the $q$'s defined by $\prod_{1\leq j\leq k}(q_j-x)=
\sum_{0\leq i \leq k}(-1)^i\sigma_{k-i} x^i $, with in particular
$\sigma_0=1$ and $\sigma_k=q_1q_2...q_k$. 

The weights $w_{i,j}$ \bolw\ are easily
interpreted in terms of different Boltzmann weights for different
types of polygons involved in the discrete world-sheet construction. 
Indeed, following the result of Sect 2.4 in the case of triangles and squares,
the coefficients $w_{i,j}$ are the Boltzmann weights of the following local
configuration of $i$ lower half-edges merging into $j$ upper ones,
all attached to the same point of the time-line:
\eqn\linew{ w_{i,j} \ \ \ \leftrightarrow \ \ \ \figbox{4.5cm}{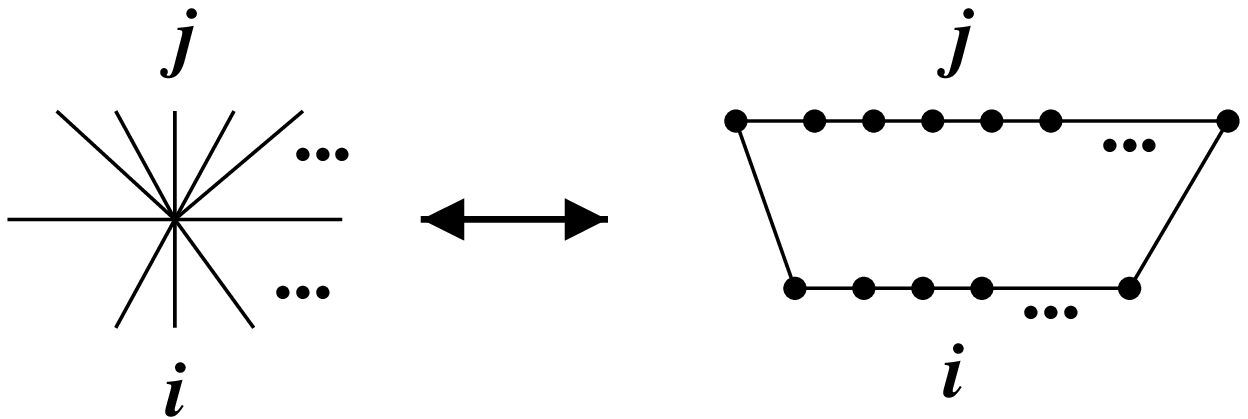} }
with the obvious dual interpretation as an $(i+j+2)$-gonal tile with two
time-like edges and $i$ lower- and $j$ upper space-like ones.
So the model whose transfer matrix is generated by \tmkq\
is one of discrete 2D Lorentzian gravity involving polygonal tiles
with 3,4,... up to $2k+2$ edges.

Note that there would be
a priori $k(k+3)/2$ independent coefficients $w_{i,j}$ in the most general
symmetric generating function of the form \tmkq, whereas here 
they are expressed in terms of only $k+1$ independent 
parameters $q_1,...,q_k,\lambda$.
Hence, we are looking at a rather small-dimensional sub-manifold of the moduli
space of 2D Lorentzian gravity, discretized with triangles, squares,
... up to $(2k+2)$-gons. The remarkable fact is that this manifold is
integrable, as the corresponding transfer matrices commute with 
one-another.
The physical meaning of the particular expression \bolw\ for the
$w_{i,j}$ weights is however not at all clear. In particular,
for more that one $q_i$, we cannot any longer map the model (as we did
in Sect.\ 2.4 for one $q$) onto a ``generalized curvature model''
with only triangles and with additional curvature weights involving 
now nearest, next-nearest,...up to, say, $k$'th-nearest neighbour 
interactions. Indeed, we can consider the most general model of triangles
with a curvature weight $a_p$ for any succession of $p$ up triangles
(or $p$ down triangles) and a weight $g$ per triangle (and as before
$\sqrt{g}$ for the boundary triangles). The integrable one-$q$
case of Sect.\ 2 corresponded to the particular choice 
$a_p=a^{p-1}$. Then is easy to 
show that the generating function $\theta(x,y\vert g,\{a_p\})$
 generalizing \generat\ reads
\eqn\gengener{\theta(x,y\vert g,\{a_p\})={1\over g}\ 
{A(gx)A(gy)\over 1-A(gx)A(gy)},}
where $A(u)\equiv \sum_{p\ge 1} a_p u^p$ is the generating function
for the weights $a_p$ (for the one-$q$ case, we simply had
$A(u)=u/(1-au)$). The particular form \gengener\ above then implies
that the quantity 
\eqn\tobefact{{g\theta(x,y\vert g,\{a_p\})\over 1+g\theta(x,y\vert g,\{a_p\})}
=A(gx) A(gy),}
must be factorized into a function of $x$ times the same function of $y$
for some particular $g$. It is easy to check that for more than one $q_i$,
the generating function $\theta_{q_1,...,q_k}(x,y\vert \lambda)$ given
by \tmkq\ does not satisfy this factorization requirement.

Before we conclude this section, let us slightly extend the definition 
\fuin\ to also include
complex (and now non-generic) numbers $q_1,...,q_k$ inside the unit disc, 
but such that $\phi_m$ remains real. A typical example would be to
take $q_1=q$, $q_2=\omega q$, $q_3=\omega^2 q$,..., $q_k=\omega^{k-1} q$,
where $\omega$ is a primitive $k$-th root of unity. Then we simply have
that $\phi_m(x\vert q_1,...,q_k)\propto F_m(x^k\vert q^k)/x^{k-1}$
with $F_m$ as in \genevec. It is clear
that the $\phi_m$'s only generate a fraction (roughly $1/k$) of the 
infinite space $E$, but the content of the model is exactly the same as 
that of the one-$q$ one. Another more interesting and possibly generic
possibility would be to include both the
$q$'s and their complex conjugates in the list, namely 
take \fuin\ with the $2k$ $q$'s being $q_1,...,q_k,{\bar q}_1,...,
{\bar q}_k$. Note that the number of free real parameters in that case would
still be $2k$. We shall make use of this extension in the next section.

Finally, we can extend the above construction so as to include other
boundary conditions. The case of periodic boundary conditions is
discussed in detail in Appendix A and corresponds to transfer matrices
which are {\it non-symmetric}. In this case, the construction requires
the use of two different (left and right) orthonormal bases.
 
\subsec{Continuum limits}

Let us now study the continuum limit of the new models (with $k$ $q$'s) 
defined in Sect. 4.2. 
In this section, we are going to derive the rather negative result that no new
scaling behaviour can be obtained from the models of Sect. 4.2, namely that 
the only well behaved continuum limit leads to the same result as in 
the one-$q$ case of Sect. 2.6.

To show this, let us first reexamine the case of one $q$.
We start from the generating
function \intcan\-\takef\ iterated $T$ times:
\eqn\genefp{ \theta_{q,T}(x,y\vert \lambda)
= (f_q(\lambda))^T {xy (1-q^2) \over (1-qx)(1-qy)-\lambda^T(q-x)(q-y)}, }  
with 
\eqn\fq{f_q(\lambda)=g{1-\lambda q^2\over 1-q^2}\ .}
To get a reasonable continuum limit, after taking $T=\tau/\alpha$ for some 
small parameter $\alpha$, we need that both $\lambda= 1+{\cal O}(\alpha)$ 
and $f_q(\lambda)= 1+{\cal O}(\alpha)$, while $g$ tends to some critical 
value $g_c<1$ when $\alpha\to 0$. From \fq\ above, this is only possible 
if $q^2=1+{\cal O}(\alpha)$ too. So we assume that $q$ tends to $1^-$ as 
$q=e^{-\alpha\sqrt{\Lambda}}=1-\alpha \sqrt{\Lambda}+{\cal O}(\alpha^2)$.
Then from $f_q(\lambda)=1+{\cal O}(\alpha)$, we deduce at leading order 
in $\alpha$ that
\eqn\dedut{\lim_{\alpha\to 0} {g_c(1-\lambda q^2)\over 1-q^2}=1
\ \ \ \Rightarrow \ \ \  \lambda= 1 -2\sqrt{\Lambda} {1-g_c\over g_c}\alpha +
{\cal O}(\alpha^2). }
Once substituted back into the generating function $\theta_q(x,y\vert \lambda)$ 
this gives 
\eqn\subtet{ \lim_{\alpha\to 0} \theta_q(x,y\vert \lambda)={ g_c xy \over 
1-(1-g_c)(x+y) +(1-2g_c) xy}, }
where we have identified the limiting Boltzmann weights \bolw\  in \tmkq\ as
\eqn\limibol{ w_{0,0}\to 1, \qquad w_{0,1}=w_{1,0}\to 1-g_c, \qquad w_{1,1}
\to -(1-2g_c),}
which remain all finite in this limit.

Let us now use the same strategy to deal with the $k$-$q$ model's 
continuum limit.
The $T$ times iterated generating function now reads
\eqn\ttim{ \theta_{q_1,...,q_k;T}(x,y\vert \lambda)=
{(f_{q_1,...,q_k}(\lambda))^T \ xy N^2(q_1,...,q_k) \over 
\prod_{i=1}^k (1-q_ix)(1-q_iy)
-\lambda^T \prod_{i=1}^k (q_i-x)(q_i-y) }  , }
with 
\eqn\fqk{f_{q_1,...,q_k}(\lambda)=g{1-\lambda(q_1...q_k)^2\over 
N^2(q_1,..,q_k)},} 
as in \takef\ and $N(q_1,...,q_k)$ as in \norma .
Setting again $T=\tau/\alpha$ and letting $\alpha\to 0$ (and 
$g\to g_c$, some critical activity per triangle), we must again have 
both $f_{q_1,...,q_k}(\lambda)=1+{\cal O}(\alpha)$
and $\lambda=1+{\cal O}(\alpha)$. Let us moreover impose that when 
$\alpha\to 0$, all the Boltzmann weights $w_{i,j}$ of \bolw\
remain finite. We must then take $N^2(q_1,...,q_k)\to 0$ and therefore 
$(q_1...q_k)^2\to 1$ to be compatible with Eq. \fqk\ above. 
But since all the $q$'s are real and less than $1$ in absolute value,
this implies the all $q_i^2 \to 1$ as $\alpha \to 0$.
Hence the natural generalization of $q=e^{-\alpha\sqrt{\Lambda}}$ to 
the $k$-$q$ case reads
\eqn\genat{ q_i=\epsilon_i e^{-\alpha_i} , \ \ 
i=1,2,...,k,}
for some small parameters $\alpha_i>0$ and some signs $\epsilon_i=\pm 1$.
If we now do not impose any longer that the $q_i$'s are real, we can
also choose pairs of conjugate complex numbers 
\eqn\genatcomp{ q_i=\omega_i e^{-\alpha_i}\ ,\ {\bar q}_i
={\bar \omega}_i e^{-\alpha_i}, }
with $\omega_i$ of modulus $1$ and such that $\omega_i^2\ne 1$.

Our choice is however strongly constrained if we insist in
having finite Boltzman weights and it can be checked that,
assuming that all the $\alpha_i$'s are of the same order $\alpha$,
this implies that the norm $N^2(q_1,...,q_k)$ itself must be of 
order $\alpha$. From the formula \othnor , we see for $k=2$ and $k=3$ that this
in practice imposes that we can have {\it at most} one real $q_i$
tending to $+1$ (i.e. $\epsilon_i=+1$ in \genat\ above) and
at most one real $q_i$ tending to $-1$ (i.e. $\epsilon_i=-1$).
Finally, to reach a non trivial limit by setting as usual
$x=1-\alpha X$ and $y=1-\alpha Y$, we need {\it at least} one  
$q_i$ tending to $+1$. We believe that these requirements are
generic for arbitrary $k$. For $k=2$, these constraints limit our
choice to exactly two real $q_i$'s as in \genat\ with 
$\epsilon_1=+1$ and $\epsilon_2=-1$.
Writing
\eqn\alwri{\alpha_i=\alpha\sqrt{\Lambda_i},\ \ i=1,2,}
we may easily compute the scaling limit of \ttim\ in  
this case $k=2$, with $x=1-\alpha X$ and $y=1-\alpha Y$, and
we get
\eqn\limty{ \lim_{\alpha\to 0}\alpha \theta_{q_1,q_2;T}(x,y\vert \lambda)=
{ C \over (X+\sqrt{\Lambda_1})(Y+\sqrt{\Lambda_1})-\phi (X-\sqrt{\Lambda_1})
(Y-\sqrt{\Lambda_1})},}
where $\phi=\lim \lambda^T$, and $C$ is a pre-factor independent of $X$ and $Y$. 
This is simply proportional to the one-$q$ result \contet, and  
corresponds therefore to the {\it same} critical behaviour.
As mentioned above, the choice
\eqn\limiom{ q_1=\omega e^{-\alpha \sqrt{\Lambda}}, \qquad q_2={\bar \omega}
e^{-\alpha \sqrt{\Lambda}}\ , }
with $\omega$ a complex number of modulus 1 and 
such that $\omega^2\neq 1$, leads to a trivial limit. Indeed, in this case, 
$N^2\sim \vert 1-\omega^2\vert^2 \sqrt{\Lambda}\alpha$,
and all the Boltzmann weights remain finite, but
in the scaling limit where $x=1-\alpha X$ and $y=1-\alpha Y$, 
all dependence in $X$ and $Y$ disappears. 
In fact, as mentioned before, to retain a dependence on $X$ and $Y$ in the 
scaling limit, we must have at least one $q$ tending to $1$, which is not 
the case here.

Applying now our constraints to the case $k=3$, we need to choose
exactly one of the $q_i$'s tending to $+1$ and the two others
being complex conjugates, i.e.
\eqn\accom{ q_1=e^{-\alpha_1}, \qquad q_2=\omega e^{-\alpha_2}, \qquad 
q_3={\bar \omega}  e^{-\alpha_2},}
with again $|\omega|=1$ and $\omega^2\neq 1$.
The normalization factor now becomes
\eqn\nortwof{ N^2(q_1,q_2,q_3)\sim {2 |(1-\omega)(1-\omega^2)|^2 
\alpha_1\alpha_2\over 
2(\alpha_1+\alpha_2)+(\omega+{\bar \omega})\alpha_2},}
and we get finite limiting Boltzmann factors by taking
\eqn\ansad{ \alpha_i= \alpha \sqrt{\Lambda_i} , \ \ i=1,2.}
Now the scaling limit of \ttim\ only retains the factors containing 
$q_1$, and wipes the others
out, so we finally get again the form \limty, with possibly 
different constants $C$ and $\phi$.

We are now ready to treat the general case. 
Depending on whether $k$ is odd or even, we must take say 
\eqn\evod{\eqalign{
k\ {\rm odd}:\ \ \ & q_1=e^{-\alpha_1}, \ q_{2i}=\omega_i 
e^{-\alpha_{i+1}}, \ q_{2i+1}=
{\bar \omega}_i  e^{-\alpha_{i+1}} \cr
&\ \ \  {\rm for}\ i=1,2,...,(k-1)/2, \cr
k\ {\rm even}:& \ \ \ q_1=e^{-\alpha_1}, \ q_2=-e^{-\alpha_2}, \ q_{2i+1}
=\omega_i e^{-\alpha_{i+2}}, \ q_{2i+2}=
{\bar \omega}_i  e^{-\alpha_{i+2}} \cr
&\ \ \  {\rm for}\ i=1,2,...,(k-2)/2, \cr}}
for some generic complex numbers $\omega_i$ of modulus 1, and 
such that $\omega_i^2\neq 1$.
As only one $q$ tends to 1, we are again led to a scaling limit of 
the form \limty. 
Note that in the end of Sect. 4.2 we have already considered the case 
when $q_j=\omega^{j-1} q$, for
$j=1,..,k$ and $\omega$ a primitive $k$-th root of unity, 
and concluded then that 
the corresponding model was equivalent to that of $k=1$, $q_1=q$ up 
to some minor rescalings ($x\to x^k, \ y\to y^k$, which do not affect 
the scaling limit): this is in agreement with the present result, 
for a particular (non-generic) choice 
of the $\omega_i$'s as $k$-th roots of unity.

To conclude, all our models have yielded the same and only non-trivial 
scaling limit
for the two-loop correlator as the one-$q$ one of Sect. 2.6. 
This is a manifestation of the rigidity of our integrability condition, 
in that all these integrable models share the same universality class.

\newsec{Conclusion}

We have revealed an interesting integrability structure underlying the
simpler models of two-dimensional Lorentzian quantum gravity and
described how one can construct various extensions of these models
while keeping them integrable. 
Among the models considered
so far, we have not found any for which the continuum properties of
the Lorentzian geometries were different from those of the pure
Lorentzian gravity case. 
We have concentrated on a model of Lorentzian quantum gravity
involving a higher curvature term which is equivalent to a model of
Lorentzian quantum gravity interacting with a simple dimer field. The
fact that this model has the same continuum behaviour as pure
Lorentzian quantum gravity is in contrast with the situation in
Euclidean quantum gravity where the introduction of dimers changes the
continuum behaviour of the geometrical system [\xref\GK,\xref\Stau].
On the other hand our
results are in compliance with the results of \AAL, where no
interaction was seen when Lorentzian quantum gravity was coupled to
an Ising spin
system. However, one should bear in mind that when
considered as a model describing matter coupled to Lorentzian gravity,
our model contains only a subset of the possible matter
configurations. It is our hope that our further investigations of the
integrability structure revealed will enable us to solve exactly more
realistic models of Lorentzian quantum gravity coupled to matter.
In the light of the recent results of numerical simulations \Apriv\
it would be particularly interesting to study the case of more than
two Ising models coupled to Lorentzian quantum gravity.
Interpreted as a model involving a higher curvature term our model is
similar in spirit to the models considered in \Kaz, where the effect of
adding a higher curvature term to usual Euclidean quantum gravity was
investigated. The result is the same for Euclidean as for Lorentzian
triangulations. Adding a higher curvature term does not change the
continuum physics for the geometrical system.

We have furthermore proven the equivalence between Lorentzian
triangulations and a certain type of random walks. This equivalence
has allowed us to set up a dictionary connecting concepts in
Lorentzian quantum gravity to concepts in the theory of random
walks. For instance, the loop-loop correlator of pure Lorentzian
quantum gravity in the language of random walks became the generating
function for large excursions. Furthermore, the integrability
structure underlying the triangle-square model of Lorentzian quantum
gravity could be understood using a simple block decomposition of the
corresponding
random walk. Finally, the  equivalence between Lorentzian
triangulations and random walks 
has provided us with an explanation why Lorentzian triangulations have
fractal dimension two and why the continuum properties of the
triangle-square model are the same as those of pure Lorentzian quantum
gravity. It is very likely that a further investigation of the
random walk picture will likewise tell us under which circumstances, 
if at all, we can expect new continuum behaviour to occur. We 
are also convinced that the above mentioned block decomposition will
prove useful in revealing the integrability structure underlying the
more realistic models of quantum gravity coupled to matter.

\vskip 0.7cm

{\bf Acknowledgements: }

P.\ D.\ F.\ thanks the organizers of the semester "Random Matrices and 
Applications"
held at M.S.R.I., Berkeley, for hospitality during the last stage of this
work. C. K.\ thanks Jan Ambj\o rn for useful discussions.

\appendix{A}{Non-symmetric transfer matrices and periodic boundary conditions}

It is instructive to derive the transfer matrix of the curvature model with
periodic boundary conditions. Its features will allow us to enhance 
dramatically the construction of Sect. 4.2. 
The transfer matrix for the periodic model
with curvature has a marked lower edge (that we will always represent as the
leftmost one), and always at least one upper edge, to avoid degeneration
into the vacuum. The transfer matrix element $T_{i,j}^{per}(g,a)$ between a row
of $i$ lower half-edges and $j$ upper ones reads
\eqn\pertran{\eqalign{
T_{i,j}^{per}(g,a)&= \sum_{k\geq 1}\sum_{\Sigma n_r=i, n_1,..,n_k\geq 1\atop
\Sigma m_r= j, m_1,..,m_{k-1}\geq 1,m_k\geq 0}
\figbox{5.cm}{tm.eps} \cr
&=\sum_{k\geq 1}\sum_{\Sigma n_r=i, n_1,..,n_k\geq 1\atop
\Sigma m_r= j, m_1,..,m_{k-1}\geq 1,m_k\geq 0} g^{i+j}
a^{\Sigma(n_r-1)+\Sigma(m_r-1)+2\delta_{m_k,0}} \cr
&=(ga)^{i+j} \sum_{k\geq 1} a^{-2k} {i-1 \choose k-1}\big[ {j-1 \choose
k-1}+a^2 {j-1 \choose k-2} \big], \cr}}
where the boundary condition is taken into account by the fact that when the
rightmost half-edge is a lower one, it receives a weight $a$ as it is the
neighbour of the leftmost one, instead of the $a^{-1}$ given by the generic
formula $a^{m_k-1}$ at $m_k=0$. The generating function for \pertran\ 
is readily found to be
\eqn\genperf{\eqalign{
\theta^{per}(x,y\vert g,a)&= \sum_{i,j\geq 1} x^i y^j 
T_{i,j}^{per}(g,a)\cr
&={g^2 x y \over (1-ga x)(1-ga(x+y)-g^2(1-a^2)xy) }.\cr} }
Let us perform the change of parameters $g,a\to \lambda,q$ of \quadrat\
and \poval. The generating function \genperf\ is transformed into
\eqn\transper{\eqalign{
\theta^{per}_q(x,y\vert \lambda)&= { g^2(1-\lambda
q^2)^2 xy \over ((1-qx)-\lambda q(q-x))
((1-qx)(1-qy)-\lambda (q-x)(q-y)) } \cr
&=\lambda\sum_{r,s\geq 1}
(1-q^2)^2{q^{r-1} x (q-x)^{r+s-2}\over (1-qx)^{r+s}}\lambda^{r+s-2} 
{y (q-y)^{s-1}\over (1-qy)^s}\cr
&= \lambda \sum_{m\geq 1} P_m(x\vert q) 
\lambda^{m-1} Q_m(y\vert q). \cr}}
where 
\eqn\lrevec{ \eqalign{ P_m(x\vert q)&= \sqrt{1-q^2} \ 
{x(q-x)^{m-1}\over (1-qx)^{m+1}}, \cr
Q_m(y\vert q)&= (1-q^2)^{3\over 2} \sum_{r,s\geq 1\atop
r+s=m+1} q^{r-1}{y(q-y)^{s-1}\over (1-qy)^s}
=\sqrt{1-q^2}\ \big(q^m -\bigg({q-y \over 1-qy}\bigg)^m\big).\cr}}
We know already that the periodic case with curvature is solvable, as we have 
computed its two-loop correlator above. The reason why is extremely simple:
the functions 
\eqn\gevecto{\eqalign{P_m(x\vert q)&=\sum_{i\geq 1} x^i {\bar w}_i^{(m)}, \cr
Q_m(y\vert q)&=\sum_{j\geq 1} y^j w_j^{(m)}, \cr}}
generate the components of two different left and right
vectors ${\bar w}^{(m)}$, $w^{(m)}$ for $T^{per}$, that turn out to be
orthonormal to one another, i.e. $w^{(m)}\cdot {\bar w}^{(m')}=\delta_{m,m'}$.
It is indeed easy to show that the contour integral over the unit circle:
$\oint P_m(x\vert q) Q_{m'}(1/x\vert q) dx/(2i\pi x) =\delta_{m,m'}$,
by use of Cauchy's residue formula.

Note however that these are not eigenvectors of $T^{per}$ but we have the
``diagonal'' mapping 
\eqn\diagmap{T^{per} {\bar w}^{(m)}= \lambda^m w^{(m)}.}
This translates immediately into the following transfer matrix relation
(as before we define $T^{per}_q(\lambda)\equiv T^{per}(g,a)$ with $g,a\to
\lambda,q$)
\eqn\transrela{ T^{per}_q(\lambda) T^{per}_q(\lambda')=
T_q^{per}(\lambda \lambda'). }
Note that this relation is also satisfied by $T_q(\lambda)\otimes
T_q(\lambda)$, thanks to \addi, and justifies a posteriori our
previous calculation of the two-loop
periodic  correlator ($p=2$ in \resop). Note finally that the two
sets of (left and right) vectors span the same space $E$, as 
$Q_m(y)\propto y$ for small $y$, and there is a triangular change of basis
from the $F_m$'s to the $P_m$'s or $Q_m$'s.

Let us draw some general conclusion from this example. 
A score of other models can be constructed by means of two different sets of
left and right vectors, provided those are mutually orthonormal. 
Given any such pair generated by say $P_m(x)$ and $Q_m(y)$, we get 
a set of transfer matrices $T(\lambda)$ with spectral parameter $\lambda$,
generated by
\eqn\genegen{ \theta(x,y\vert \lambda)=\sum_{i,j\geq 1}
T_{i,j}(\lambda) x^i y^j= f(\lambda)\sum_{m\geq 1} P_m(x) \lambda^{m-1}
Q_m(y), } 
for some arbitrarily chosen function $f(\lambda)$. The vector spaces
spanned by the left and right vectors need not be the same, we simply need the
vectors to be mutually orthogonal: $\oint dx/(2i\pi x) P_m(x)
Q_{m'}(1/x)=\delta_{m,m'}$. This implies a multiplicativity relation
\eqn\multipl{ T(\lambda) T(\lambda')= {f(\lambda)f(\lambda')\over
f(\lambda\lambda')} T(\lambda\lambda').}
which makes the model trivially integrable.

In the light of this, it is easy to find a non-symmetric 
generalization of the $k$ $q$'s model of Sect. 4.2. Taking
\eqn\gekqf{\eqalign{
P_m(x)&= M_k\ x\ \prod_{j=1}^k {(q_j-x)^{m-1} \over (1-q_jx)^{m+1}}.
\cr
Q_m(y)&= (-1)^{k-1}M_k \bigg( (q_1q_2...q_k)^m -\prod_{j=1}^k 
{(q_j-y)^m \over (1-q_jy)^m} \bigg).\cr}}
as generating functions of the left and right vectors respectively,
it is easy to prove that these are mutually orthonormal, provided
we take
\eqn\norM{\eqalign{
{1\over M_k^2}&=(-1)^k\oint {dx \over 2i\pi} \prod_{j=1}^k{1\over 
(1-q_jx)(q_j-x)} \cr
&=\sum_{j=1}^k \prod_{i=1}^k{1 \over 1-q_iq_j}
\prod_{i=1\atop i\neq j}^k{1 \over q_j-q_i}.\cr}}
by Cauchy's formula. 
For $k=1$, this gives $M_1=1-q^2$, in agreement with \lrevec, while
for $k=2,3$ we have
\eqn\ktwothree{\eqalign{
&M_2^2= {(1-q_1^2)(1-q_2^2)(1-q_1q_2) \over q_1+q_2}. \cr
&M_3^2= \cr
&{(1-q_1^2)(1-q_2^2)(1-q_3^2)(1-q_1q_2)(1-q_2q_3)(1-q_1q_3)\over
q_1^2+q_2^2+q_3^2+q_1q_2+q_2q_3+q_1q_3-q_1q_2q_3(q_1+q_2+q_3)-
q_1^2q_2^2-q_2^2q_3^2-q_1^2q_3^2}.\cr}}
The vectors \gekqf\ lead to the transfer matrix generated by
\genegen\
\eqn\otet{\eqalign{
\theta(x,y\vert \lambda)&=
xg(\lambda){\prod_{j=1}^k q_j(1-q_j y)-\prod_{j=1}^k (q_j-y)\over
\prod_{j=1}^k(1-q_jx)-\lambda\prod_{j=1}^k q_j(q_j-x)} \times \cr
&\qquad \times {1 \over \prod_{j=1}^k(1-q_jx)(1-q_jy)-
\lambda \prod_{j=1}^k (q_j-x)(q_j-y)} \cr}}
where $g(\lambda)$ is an arbitrary function of $\lambda$ (in which we have 
absorbed various normalization factors).
We suspect the transfer matrix of \otet\ to be the periodic
boundary condition version of \intcan\-\takef.
Note that here the vectors generated by \gekqf\ both span the
same space $E$ (they are proportional to $x$ and $y$ respectively).

Let us finally mention the following generalization of \gekqf, based
on the general property that if $P_m(x)$ and $Q_m(y)$ are
generating functions of mutually orthonormal vectors, then
for any function $R(x)$, the vectors generated by 
$\phi_m(x)=R(x) P_m(x)$ and $\psi_m(y)=Q_m(y)/R(1/y)$ are also
mutually orthonormal (we must of course assume that both
$R(x)$ and $1/R(1/y)$ have convergent series expansions). 
We may apply this recipe to \gekqf\ with the function
$R(x)=1/(\prod_{j=1}^k (1-q_jx)^{p_j-1})$, for some
given integers $p_j\geq 1$. But 
$1/R(1/y)= (-1)^{p-k} \prod_{j=1}^k (q_j-y)^{p_j-1}/y^{p-k}$,
where we have defined $p=\sum p_j$, 
has no good series expansion around $y=0$. In order to fix this and 
arrange for both the left and right vectors to generate the same
space $E$ as above, we are led to take
\eqn\phipsi{\eqalign{
\phi_m(x)&= {S_k} x \prod_{j=1}^k {(q_j-x)^{m-1}\over
(1-q_j x)^{m+p_j}} \cr
\psi_m(y)&=(-1)^{p} {S_k} {1\over y^{p-k}}\bigg(
\prod_{j=1}^k {(q_j-y)^{m+p_j-1} \over (1-q_j y)^m} -
\sum_{l=0}^{p-k} \alpha_l y^l \bigg) \ .\cr}}
The coefficients $\alpha_l$ are fixed 
by requiring that 
$\psi_m(y)=O(y)$, and $S_k$ is the normalization factor ensuring orthonormality,
with the result $S_k=M_k$ defined in \norM.   
Note that for any polynomial $P$,  $P(1/y)$ is
automatically orthogonal to $\phi_m(x)$, as no pole lies inside the unit
disc in the corresponding Cauchy integral 
$\oint \phi_m(x) P(x) dx/(2i\pi x)=0$ for all $m\geq 1$
(the $x$ in the denominator is cancelled by the one in factor of $\phi_m$).

\listrefs
\bye